\documentclass[12pt,preprint]{aastex}
\usepackage{longtable}

\makeatletter
\def\jnl@aj{AJ}
\ifx\revtex@jnl\jnl@aj\let\tablebreak=\nl\fi
\makeatother

\slugcomment{Submitted to \aj}

\shorttitle{Variable stars in COROT IR01 field}

\shortauthors{Kabath et al.}

\begin{document}

\title{\begin{center}Characterization of CoRoT target fields with BEST: Identification of periodic variable stars in the IR01 field\end{center}}

\author{Kabath P., Eigm\"uller P., Erikson A., Hedelt P., Rauer H.\altaffilmark{1}, Titz R., Wiese T.}

\affil{Institut f\"ur Planetenforschung, Deutsches Zentrum f\"ur Luft- und Raumfahrt, 12489 Berlin, Germany}

\and

\author{Karoff C.}

\affil{Department of Physics and Astronomy, University of Aarhus, Ny
Munkegade, Building 1520, DK-8000 Aarhus C, Denmark}

\email{petr.kabath@dlr.de}

\altaffiltext{1}{Also: Zentrum
f\"ur Astronomie und Astrophysik, Technische Universit\"at Berlin,
Germany}

\begin{abstract}

We report on observations of the CoRoT IR01 field with the Berlin
Exoplanet Search Telescope (BEST). BEST is a small aperture telescope with a
wide field of view (FOV). It is dedicated to search for variable
stars within the target fields of the CoRoT space mission to aid in
minimizing false-alarm rates and identify potential targets for
additional science. CoRoT's observational programm started in
February $2007$ with the "initial run" field (IR01) observed for
about two months. BEST observed this field for $12$ nights spread
over three months in winter $2006$. From the total of $30426$ stars observed in the IR01 field $3769$ were marked as suspected variable stars and $54$ from them showed clear periodicity. From these $19$ periodic stars are within the part of the
CoRoT FOV covered in our data set.

\end{abstract}


\section{Introduction}

Small ground based surveys with wide field of view (FOV) like the
Berlin Exoplanet Search Telescope (BEST) \cite{Rauer2004} can
provide useful information on the identification of variable stars
in the target fields of photometric space missions like CoRoT. The
CoRoT mission \cite{Baglin1998} performs observations of stellar
pulsations of selected stars (asteroseismology) and photometric
searches for transit events of extrasolar planets. In addition, a
programm on "additional science" studies is performed on dedicated
objects, including variables, binaries, etc. Ground based surveys
like BEST can play an important role to pre-check for stellar
variability and to classify the newly detected variable stars in the
stellar fields which will be later observed with CoRoT.

CoRoT has been successfully launched in December $2006$ and the
first scientific data are expected in early summer $2007$. After the
commissioning phase CoRoT began to observe the first short
"initial run" field (IR01) for about two months. The following mission
phases include long observational runs ($150$ days per each field)
which are complemented with shorter runs ($30$ days per
field) \cite{Michel2006}.

In addition to transits of extrasolar planets, transit shaped
photometric signals can be caused by, for example, stellar spots,
variability or eclipsing binaries. To distinguish between stellar
activity and real planetary transits, prisms were mounted in front
of the CoRoT detectors in the exoplanet channel in order to obtain
colour information on the lightcurve \cite{Barge2006}. Due to
the prism the point-spread function (PSF) covers about $80$ pixels
on the exoplanet detection part of the CoRoT
CCDs \cite{Boisnard2006}. Thus, the overlapping PSFs of stars in
crowded fields may hide the signal originating from a planetary
transit event. Another problem arises from the fact that
transit-like lightcurves may actually be false alarms due to stellar
variability. Eclipsing binary systems, especially grazing eclipses,
show transit-shaped lightcurves very similar to an extrasolar planet
transit. Therefore, it is of advantage to know the distribution of
variable stars in the CoRoT target fields prior to the spacecraft
observations. Incoming lightcurves from the spacecraft can be
checked for already known variables overlapping with the selected
PSF-window, or the information can be used to avoid such variables
when placing the aperture windows of the satellite.  For the
additional science programm, however, variables may be interesting
targets. Information on such variable stars can be provided by a
ground based survey like BEST.

BEST observes the CoRoT stellar fields in order to identify
variables and to preselect interesting objects prior to CoRoT's
observations. This information will be fed into the EXODAT
database \cite{Deleuil2006} which is a catalogue containing information on more than $10$ millions of stars based on mainly photometric observations prepared for the CoRoT observational sequence adjusting. Here we focus on variable stars in IR01
field. The BEST results on the stellar variability in the CoRoT LRc1
field (the first "long run" field of 150 days observing time) can be
found in \cite{Karoff2007}.

The paper is arranged as follows. In the following section a short
description of the BEST system is followed by a description of the
observations. The third section presents the data reduction
pipeline, including a description of photometric and astrometric
routines used. The stellar variability characterization and the
classification of the newly identified variable stars is included in
the fourth section. The last section is an overview on the presented
data.

\section{BEST setup and observations}

BEST is operated by the "Institute f\"{u}r Planetenforschung" of the "Deutsches Zentrum f\"{u}r Luft- und Raumfahrt (DLR)" and was located in Tautenburg, Germany from $2001$ until
$2004$ \cite{Rauer2004}. It was then relocated
to the Observatoire de Haute-Provence (OHP), France. BEST started to
regularly observe from OHP in early $2004$ with a maintenance period
of three months in summer $2006$ which were without observations.
Since autumn $2006$ the telescope is operated in remote control mode
by an observer from Berlin. The primary goal of the observations is to identify new variable stars in the target field
of the CoRoT mission.

BEST consists of a Schmidt-Cassegrain telescope with a $19.5$ cm
effective aperture and a focal ratio of $f/2.7$. The telescope is
equipped with a $2048\times 2048$ pixel CCD Apogee AP-$10$ with grade E chip with pixel size of $14 \times 14\mu $m. The image scale corresponds to $5.5''$ per pixel, and the FOV is $3.1^\circ
\times 3.1^\circ$. The digital resolution is $14$ bit, resulting in a saturation level of $16384$ ADU and a short readout time of $9$ sec. The quantum efficiency of the chip reaches $40\%$ for wavelengths within $650-800$ nm. Observations are performed without any filter in
order to achieve the maximum photonic gain. The spectral response of
the CCD corresponds to a wide $R$-bandpass filter. Due to the large
FOV BEST can detect more than $30000$ stars in a typical stellar
field positioned in the galactic plane.

\begin{figure}
\begin{center}
\includegraphics[angle=90, scale=.45, width=12cm]{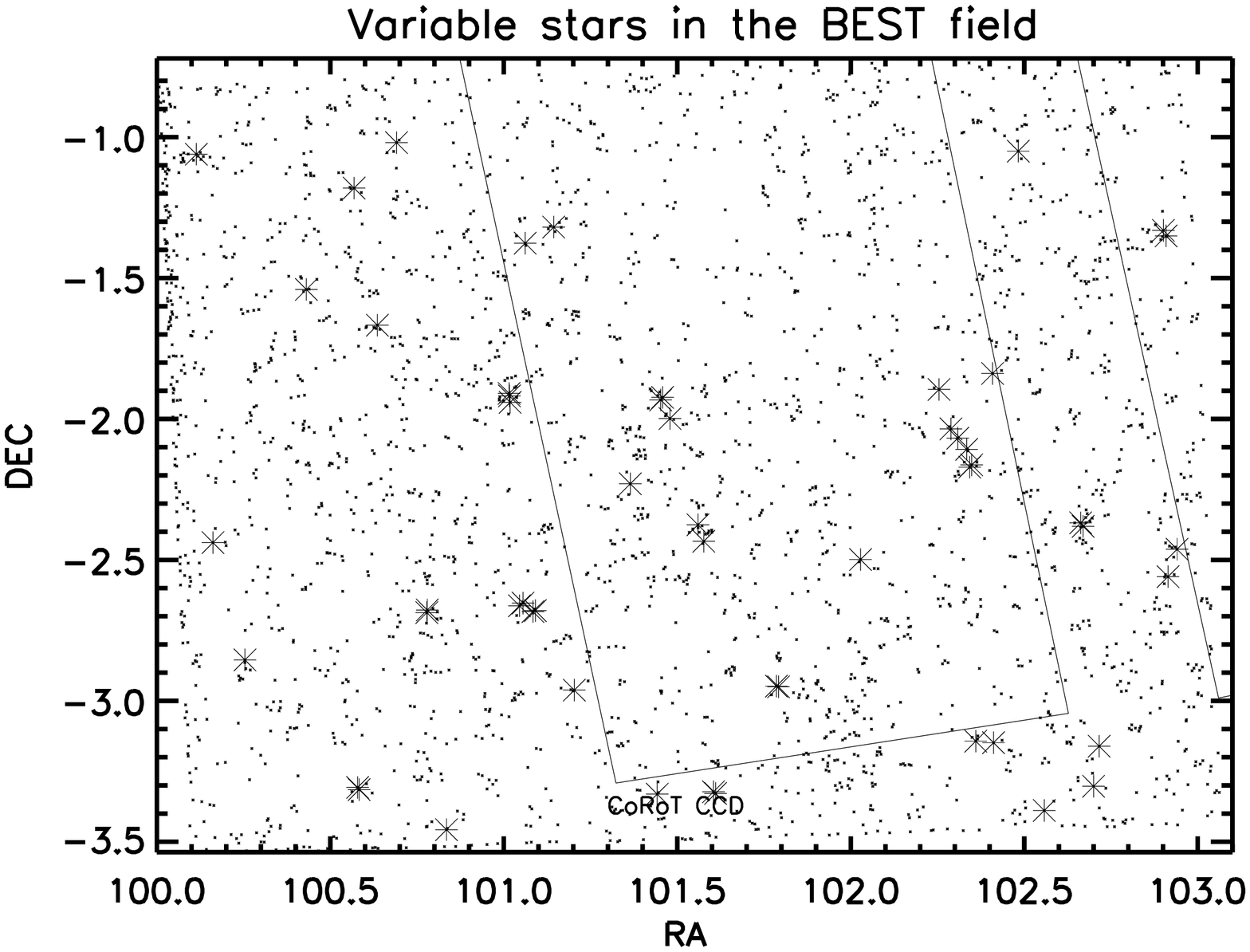}
\caption{The CoRoT IR01 field is indicated in the BEST field. The distribution of the variable stars found in the BEST FOV is shown. Dots show stars with $j>2$ and stars are periodic variables selected visualy from the stars having $j>2$. \label{cam}}
\end{center}
\end{figure}

The center of the observed IR01 field is located at $\alpha=6^h 57^m
18^s$ and $\delta=-1^\circ 42' 00'' $. The BEST FOV was oriented
such that the exoplanet part of the CoRoT FOV was covered, but the
very bright stars of the seismology part of the were
safely avoided (see Fig. \ref{cam}) - the approximate coordinates of the center of BEST FOV are $\alpha=06^h 46^{m} 24^{s}$ and $\delta=-01^\circ 54' 00''$. We observed the IR01 field from the
beginning of November, when it began to be above the horizon for
more than $3$ hours. The observing campaign finished in
mid-December. BEST collected the data from $12$ nights, spread over
the whole observational period. No data were obtained at nights with
bad weather conditions and around full Moon. We observed $5$ nights
under very good, $5$ nights under good and $2$ nights under poor
photometric conditions. In total we observed the field over the time span of $40.012$ days.

\begin{figure}
\begin{center}
\includegraphics[angle=90,scale=.40, width=12cm]{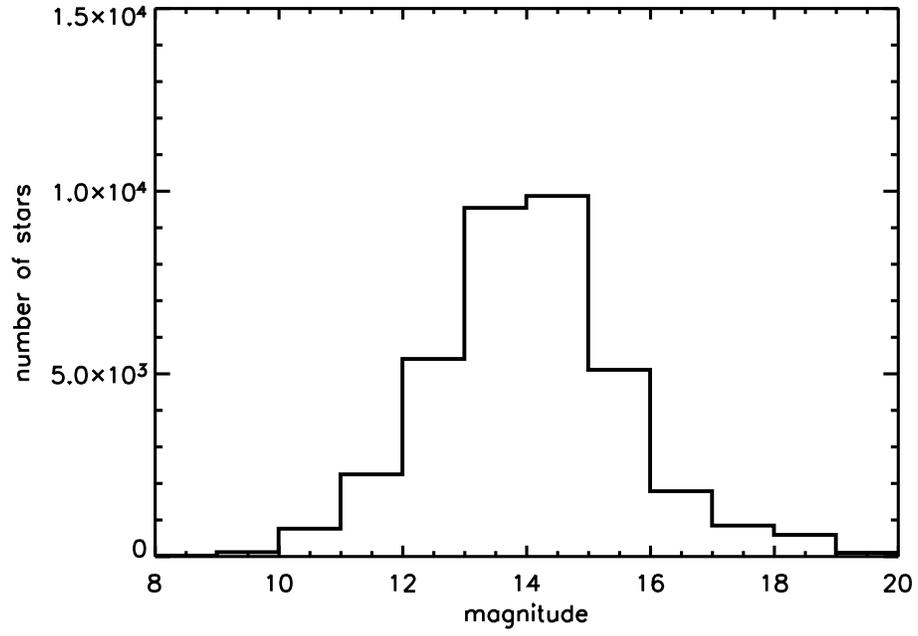}
\caption{Histogram of magnitudes of the detected
stars in the BEST FOV.\label{hist}}
\end{center}
\end{figure}

The observing run consists of images with two different exposure
times in order to cover a wide range of detectable magnitudes. A
$40$ second exposure was followed by $240$ seconds exposures. The
apparent brightness of the detected stars ranges from $8$ to $18$
magnitudes. A histogram of the stellar magnitudes of the 240 second
exposures is shown in Figure \ref{hist}. Although bright stars are
detected in the frames with long exposure time, stars with
magnitudes less than $10$ mag can be saturated, depending on seeing
and transparency, or affected by CCD non-linearities. The bright
stars therefore need to be studied in the short exposures. In the
following, we present only the data obtained from $300$ frames with longer exposure time which corresponds to total $300$ data points per lightcurve and with magnitudes $>$ $10$ mag because the faint stars overlap with
the magnitude range of CoRoT (12 - 16 mag) and according to our test the CCD is working linear at these magnitudes for $240$ seconds exposure time.   

Each exposure sequence was followed by bias and dark frames for
calibration purposes. The Apogee CCD camera is only cooled
with a $2$-stage thermoelectrical Peltier cooler with forced air. Cooling temperature of maximum $38^\circ C$ below ambient temperature can be reached but a significant dark current is detected. Variations in
temperature stabilization lead to small variations in the dark
current level which are monitored during the night. However, the variations are typicaly below $0.5\%$ during the night. Bias level variation are at the same level. The bias frames obtained during the observations are mainly used to monitor the performance of the CCD and to clean it from any local charge levels e.g. from the saturation due to bright stars.

\section{Data reduction}

\subsection{Basic calibrations and data pipeline}

The data reduction and analysis follows \cite{Karoff2007}. In the first step a basic
photometric calibration of the raw CCD data is performed, including
dark current and bias subtraction as well as flatfield correction
(for details see Rauer et al. (2004)). The dark and bias frames taken regulary within the observing sequence are checked for variations during the night and they are then composed into a master frames used for the calibration of the raw images, together with calibrated master flat fields. In order to perform a quick and
accurate reduction of the frames which also removes the sky
background, an image substraction routine from the ISIS data
reduction package \citep{Alard1998} was used. Prior to the
photometric reduction an image with the best seeing from
the middle of the observation run was chosen from the data set. This
image was used as a template for the transformation of $(x,y)$
coordinates of all frames into the same coordinate grid. After the
coordinate transformation ISIS creates a reference frame by
combination of the images with the best seeing (in the case of IR01 we used $14$ images). The reference
frame is then subtracted from all other images, such that each PSF
of a star in each frame is convolved with the kernel of the
corresponding star in the reference frame. Simultaneously the sky
background is removed.

\subsection{From calibrated frames to lightcurves}

Photometry was performed as simple unit-weight aperture photometry
on the subtracted frames and on the reference frame to obtain
instrumental magnitudes. The analyzed stars have a FWHM between $1$
and $3$ pixels, depending on seeing. Therefore we use an aperture
radius of $7$ pixels, which was found to be optimal for our
purposes \cite{Karoff2007}.

After applying the photometric routine, the brightest $5069$ stars
(magnitude limit $11.4$ mag) of the BEST data set were matched with
stars from the USNOA-2.0 catalogue \cite{Monet}. This catalog was
used as reference to obtain a rough calibration of our instrumental magnitudes and to perform astrometry. We note that our variable
search requires only relative magnitudes and we do not aim on an
accurate absolute photometric calibration of our target stars. For
the astrometric calibration, we use a cubic transformation within
the MATCH routine \cite{Valdes} to calculate the transformation from
the $(x,y)$ coordinates of the BEST-CCD to $\alpha$ and $\delta$. A
transformation of lower order did not work because of large FOV of
BEST. For the matching itself we used a radius of $2\cdot 10^{-4}$
deg. Difficulties due to multiple matches are a problem using this
procedure because of the large image scale of BEST ($5.5''$
pixel$^{-1}$). However, $2945$ stars could be successfully matched
with the catalogue stars. The derived magnitude correction was then
applied to all stars from the whole data set.

\subsection{Photometric accuracy}

BEST data can be affected by systematic errors due to zero-point
offsets from night to night or a large scatter due to bad weather
conditions. Observations of a very good night contain more than
$3000$ stars with rms scatter of the lightcurves less than $1\%$.
This precision can be reached in the magnitude range $10-13$ mag in
such good nights. However there is a clear difference between nights
with good and those with bad photometric quality which is reflected
by the varying number of stars having an rms below $1\%$. These
effects can lead to a misidentification of variable stars with
automatized selection routines. In order to minimize these and
others effects leading to remaining systematic intensity variations
in the data after application of the reduction pipeline, a cleaning
algorithm \cite{Tamuz2005} was applied four times on the whole data
set. The algorithm detects systematic intensity variations affecting
the whole data set (e.g. residual extinction) and therefore performs
an additional photometric correction to the data. This correction is
most effective for bright stars, faint stars are dominated mainly by
background noise. The resulting lightcurves from the cleaning algorithm were then searched for the variable stars.

\begin{figure}
\begin{center}
\includegraphics[angle=90, scale=.30, width=14cm]{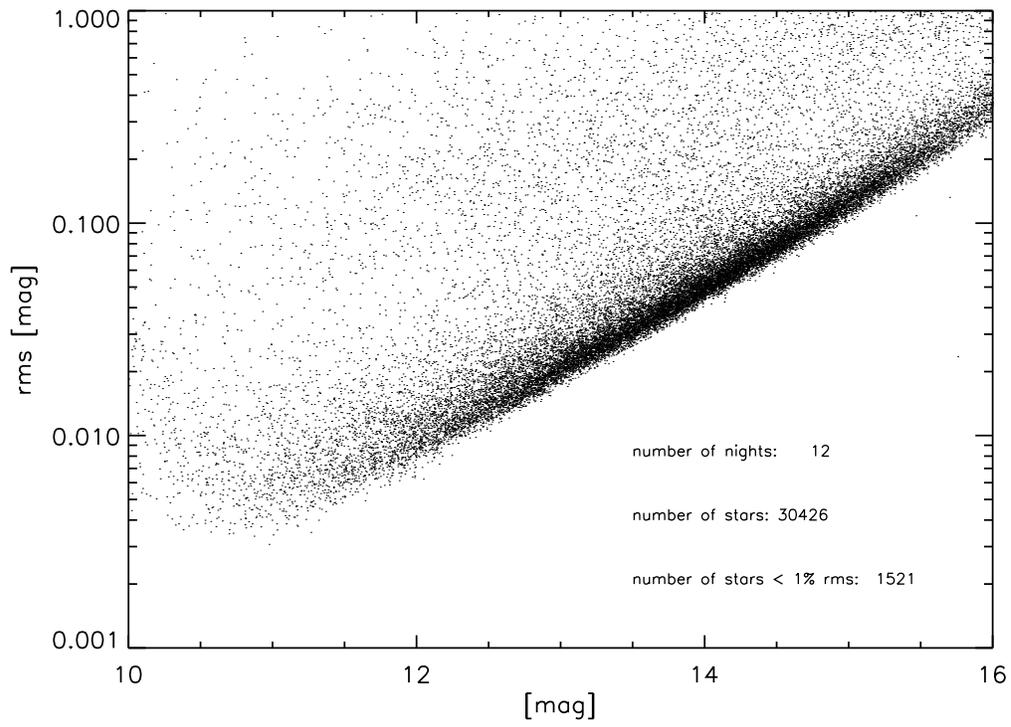}
\caption{The rms noise level of stellar lightcurves for the whole data set over magnitude over the observing
campaign.\label{rms}}
\end{center}
\end{figure}

Figure \ref{rms} shows the resulting rms of the stellar lightcurves
in our data versus magnitude over the whole observing campaign.
About $1500$ stars in the magnitude range of $10-12$ mag have an rms
level of less than $1\%$ in the combined lightcurves, clearly showing the effect of remaining poor observing nights on the overall photometric quality. We nevertheless decided to keep all nights in
the already time limited data set. The scattered very high levels of
rms seen at all magnitudes are caused in part by real intrinsic
variability of stars (see below), but can in part also be caused by
instrumental effects, such as cosmic ray hits, stars crossing hot
CCD pixels, etc., or effects by crowding and varying seeing.

The large pixel scale of BEST and the dense CoRoT field result in
reduced photometric quality in very crowded areas of the field. The
effects of crowding are already significantly reduced, however, by
using ISIS instead of simple direct aperture photometry. Photometry
is applied in reference subtracted frames with ISIS. In these frames
non-variable stars cancel out and aperture photometry of the
remaining variables can be performed with relatively large radius.
However, in very dense areas variable stars or residuals of bright
stars within the aperture of a detected star can affect the result.
This leads to enhanced noise, but can also lead to multiple
detections of the same variable. To avoid such false multiple
variable detections, each periodic variable star was checked manually to
securely identify the right object.

Variable background objects within the PSF of BEST may lead to
diluted signals or may be missed as variables altogether.
Identification of such background objects needs higher spatial
resolution follow-up observations. Thus, the possibility that
variables reported here are actually caused by background objects or
unresolved stars within our PSF can not be excluded and is subject
of future dedicated follow-up investigations on interesting objects.

\section{Variable stars}

The lightcurves resulting from our calibration pipeline and after
further correction of systematic residual intensity variations by
the algorithm of \cite{Tamuz2005} are then searched for stellar
variability. The motivation for our search for variable stars in the CoRoT field has been mentioned in the Introduction to this paper. Selection process, period search and classification of the identified variable stars is described here.   

\subsection{Criterion for variability}

\begin{figure}
\begin{center}
\includegraphics[angle=0, scale=.45, width=14cm]{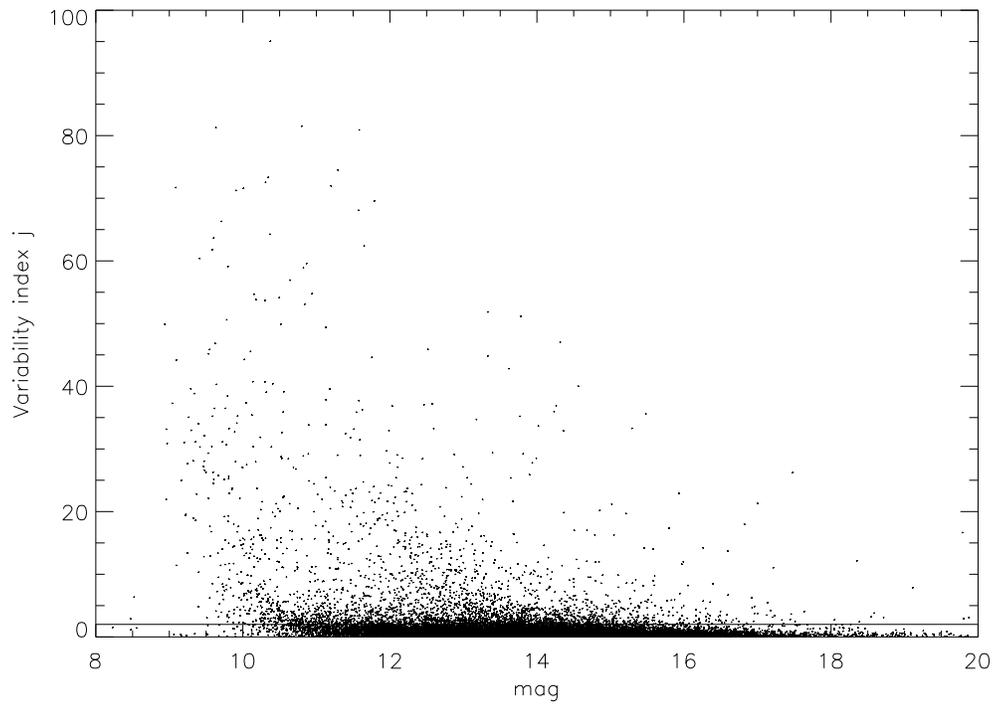}
\caption{Variability index $j$ of the sample plotted over magnitudes
of the stars. The line marks the limit of $j=2$ used to identify
suspected variable stars.\label{jindex}}
\end{center}
\end{figure}

In order to classify the activity of the stars contained in the data
set a variability index $j$ is defined as the normalized sum of the
deviations of each data point (or pair of data points with small
separations in time) in the lightcurve compared to the expected
noise level for the star. We changed the original definition of the
Stetson's variability index \cite{Stetson1996} according
to Zhang et al. (2003). We used $j>2$ as a criterion for suspected
stellar variability. Figure \ref{jindex} shows the index $j$ as a
function of magnitude. $3769$ stars were identified as possible
variables from a total number of $30426$ detected stars.
Figure \ref{cam} shows the spatial distribution of the suspected
variables in the BEST FOV.

Only for a part of the suspected variable stars which were marked with $j$ index a period was found. However, there
will be a number of additional periodic variables, in particular
long-periodic, for which periodicity can not be determined in the
sparse data set. An additional fraction of the suspected variables
will show real, but non-periodic variability. The remaining stars
with high $j$ index will show variability caused by instrumental
effects or high noise levels remaining after photometric reduction.
It is difficult to discriminate the latter from real non-periodic
variables, because instrumental effects can depend on the position
of a star on the CCD (e.g. hot pixels) or with time (e.g. cosmic ray
hits) and thus are difficult to estimate on a statistical basis. In this paper we
concentrate on periodic variable stars, because they
are of most interest for CoRoT. However, the information of
suspected variables with $j>2$ will be made available to interested
observers on request for further investigations.

\subsection{Detected periodic variables}

All lightcurves of suspected variable stars were searched for the
periodicity with the method introduced by Schwarzenberg-Czerny (1996). This method
fits a set of periodic orthogonal polynomials to the observational
data sets and evaluates the quality of the fit using an analysis of
variance statistics. All suspected variables with fit quality better
than $0.9$ were examined visually. Most rejected of suspected
variable stars show periods close to one day or multiples of one day
and show no clear evidence of at least two maxima or minima in the
lightcurve. This rejected variability is very likely caused by
changes of the background level, temperature, airmass or by instrumental effects.

In total, $54$ stars were classified as periodic variable stars
after the visual check. Periods were searched between $0.1$ day up to
$10$ days. The details about the stars are listed in
Table 1. A distribution of these newly detected periodic
variable stars over the BEST FOV and also within the CoRoT IR01
field covered can be seen in Figure \ref{cam}. The number of
variable stars detected in the CoRoT FOV covered by our
observations is $19$ and they are marked with a star in Table 1. The finding charts will be provided upon a request.

We compared the periodic variable stars identified with the GCVS
catalogue via SIMBAD coordinates query. Two BEST stars (ID $2109$
and $2390$) were already present in GCVS: $V 453$ Mon and $V 515$
Mon. In total, $52$ of our
variable stars are new discoveries.

The completeness of our data set is restricted. We have been able to
detect  only short periodic variable stars due to bad weather
conditions in mid November and in general due to the short observing
run. The IR01 field was observable from $3$ (at the beginning of
November) to a maximum of $6$ (mid December) hours per night only.
Therefore the periods of detected variable stars found are usually
below $1$ day. The following stars showed periods longer than one
day: ID $1101$, $1180$, $1194$, $1239$, $1549$, $1865$, $1993$, $2440$, $2632$, $2633$, $2651$, $2661$,
$3175$ and $3203$. Star $1194$ is located near to a bright star for which a reasonable match is the star HD295427, however it is not possible to decide which star from both of them is the variable star. Star $2349$ shows a "corrupted" lightcurve which may be due to nearby
star $2354$.

The classification of the periodic variable stars is based on the GCVS (see Sterken $\&$ Jaschek, 1996). As only a limited amount of information is available at lightcurves contained in our data set -- i.e. we do not have any spectral or color information, we have not been able to make a full classification into all the classification classes given by the GCVS. The classes we have used are: CEP, DSCT, ELL, EA, EB and EW. 

The main separation in the classification is between pulsating periodic variable stars, eclipsing binaries and rotating stars. This classification is relatively straight forward to make. The pulsating variable stars have been separated in to CEP and DSCT based on the period. Stars with period below $0.3$ days were classified as DSCT and stars with periods longer than $0.3$ days were classified as CEP stars what is based on the earlier usage of the CEP class without subdivisions. Thus CEP refers here to all kind of pulsating periodic stars with periods longer than $0.3$ days, e.g. BCEP, CEP, DCEP, RR as the information does not allow the classification into the subclasses. For the DSCT with periods below $0.3$ days the classification should be fairly unique. The classification for the rotating stars ELL is based on the unequal minima/maxima of the lightcurves.

The eclipsing binaries have been separated into EA, EB and EW classes based on the criteria given by Sterken \& Jaschek (1996) -- i.e EA class shows constant lightcurves between eclipses, EB class varies continuously between eclipses and stars belonging to the EW class have equal depth and periods shorter than 1 day.

The stars $1194$, $1898$ and $2633$ show clear variability, but no clear period. We expect that the stars are either multiperiodic pulsating stars (e.g. BCEP, SPB, $\gamma$ Dor stars) or rotating spotted stars. However more data with better duty cycle is needed to make a precise classification. Such data will be provided with CoRoT. These three stars are marked in the Table 1 with ?. 

\section{Summary}

The BEST telescope was used to search for short-period variables
within the CoRoT exoplanet channel FOV of the IR01 field. In total $54$
periodic variable stars were detected and $52$ of them are new
discoveries. From these stars, $19$ are within the part of the CoRoT
field covered in our observations. All detected variables are short
period (mainly P $<$ $1$ day), because of the limited data set.
Nevertheless, the variables found will give some additional
information when analysing CoRoT PSFs near these variables and will
help avoiding false-alarms in case of potential transit candidate
signals nearby.

\acknowledgments

The authors gratefully acknowledge the support and assistance of the staff at Observatoire de Haute-Provence. We are also most grateful to an anonymous referee for his helpful comments and useful advice.

We would like to greatly appreciate the help of Sabrina Kirste who
contributed to preparing of the manuscript. C. K. acknowledges
support from Danish AsteroSeismology Centre and Instrument Center
for Danish Astrophysics. The support to DLR staff in observational
tasks given by Berlin local amateur astronomers Susanne Hoffman,
Irene Berndt, Martin Dentel and Karsten Markus is also greatly
acknowledged. We made use of SIMBAD and GCVS catalogues.

\clearpage

\clearpage

\clearpage

\clearpage

\begin{figure}[ht]
\includegraphics[scale=.45]{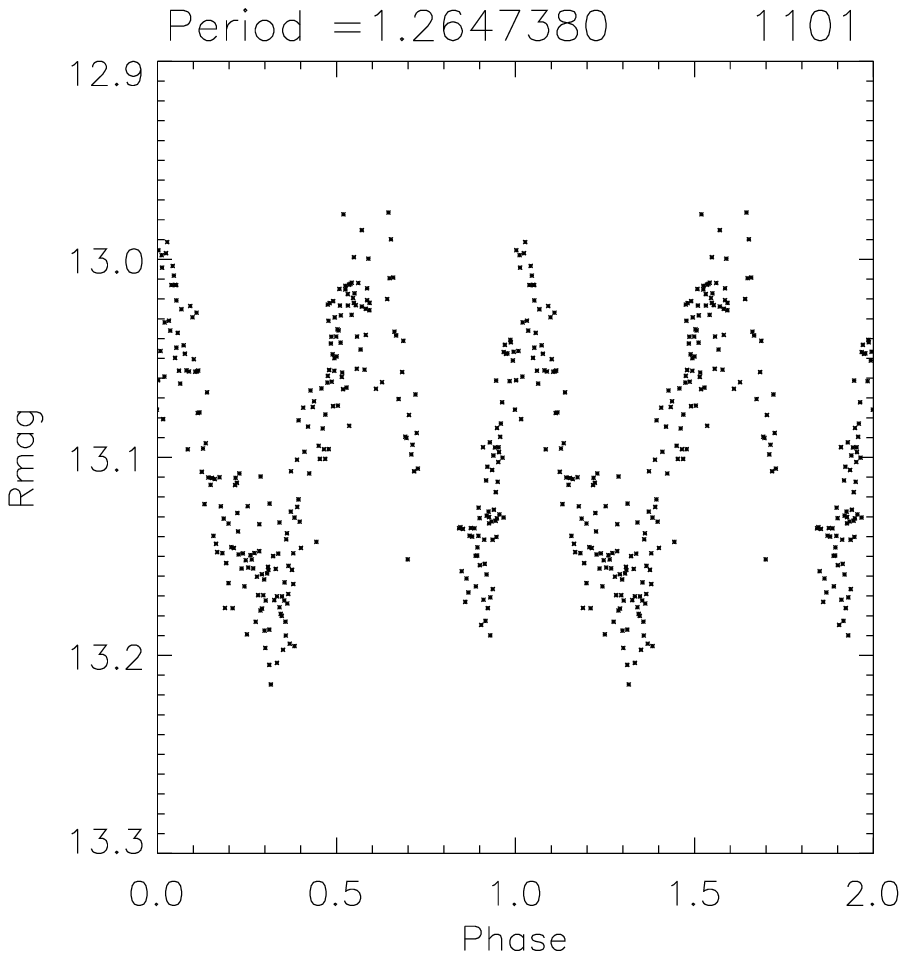}
\includegraphics[scale=.45]{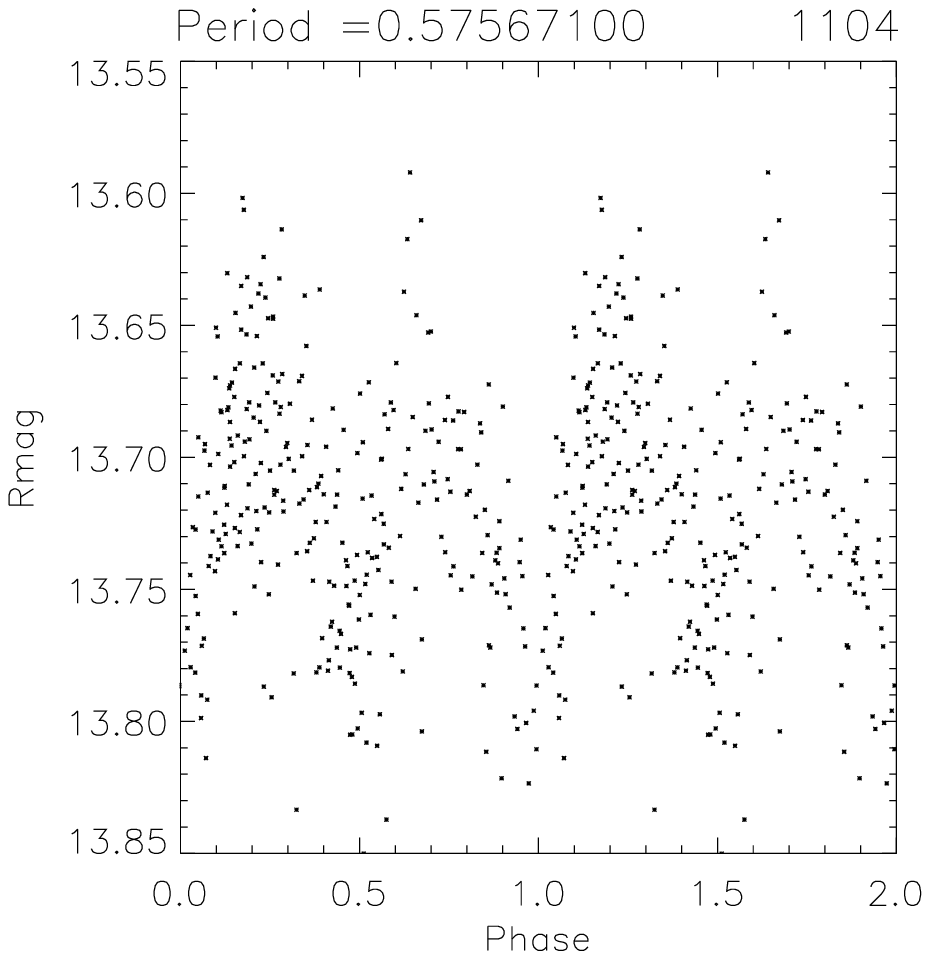}
\includegraphics[scale=.45]{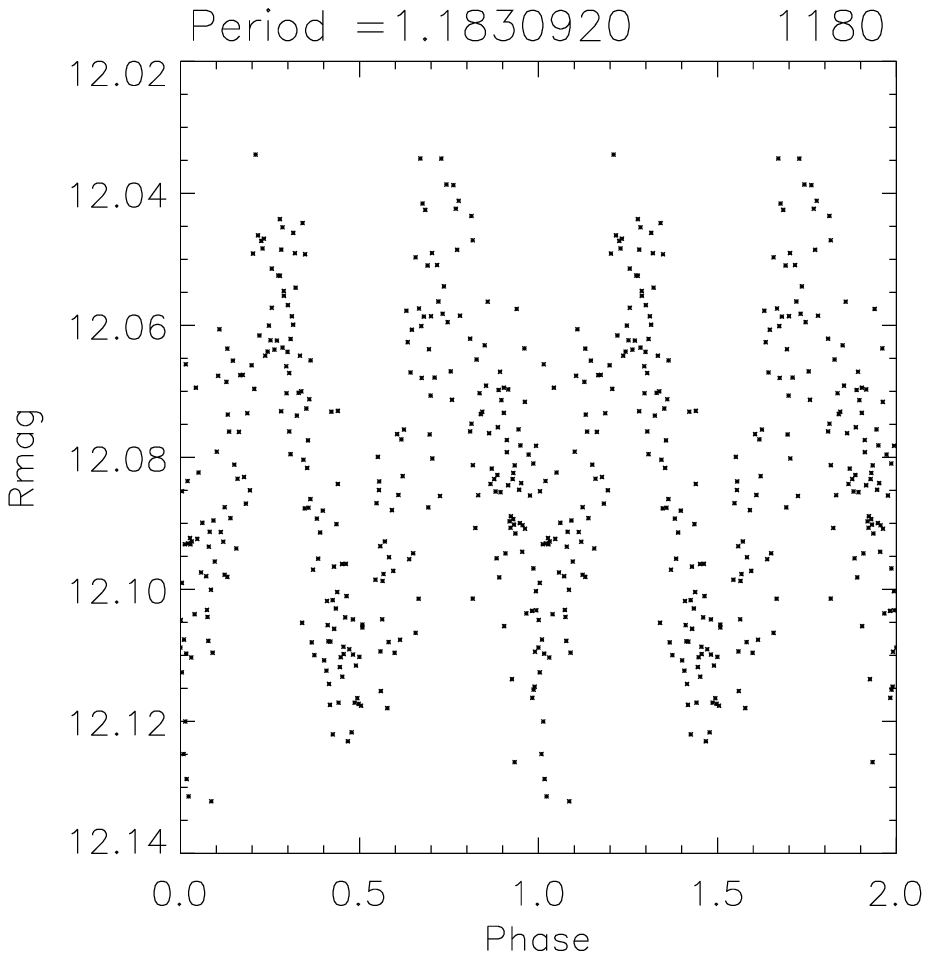}
\includegraphics[scale=.45]{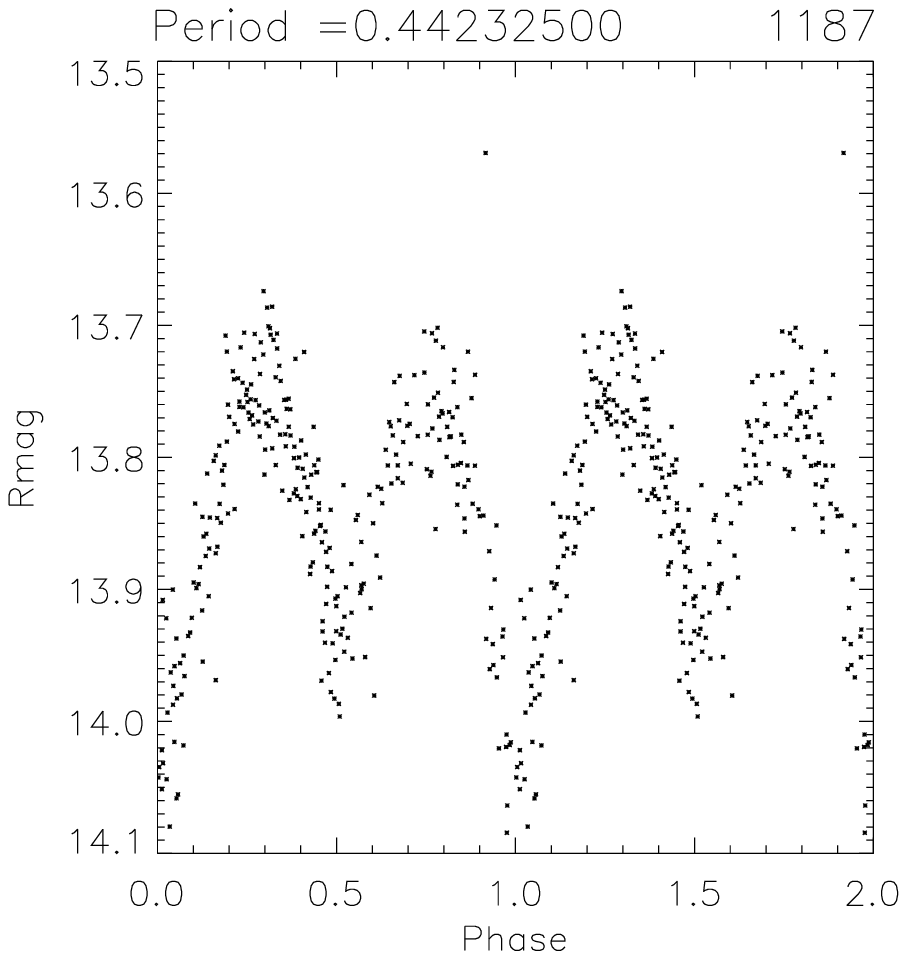}
\includegraphics[scale=.45]{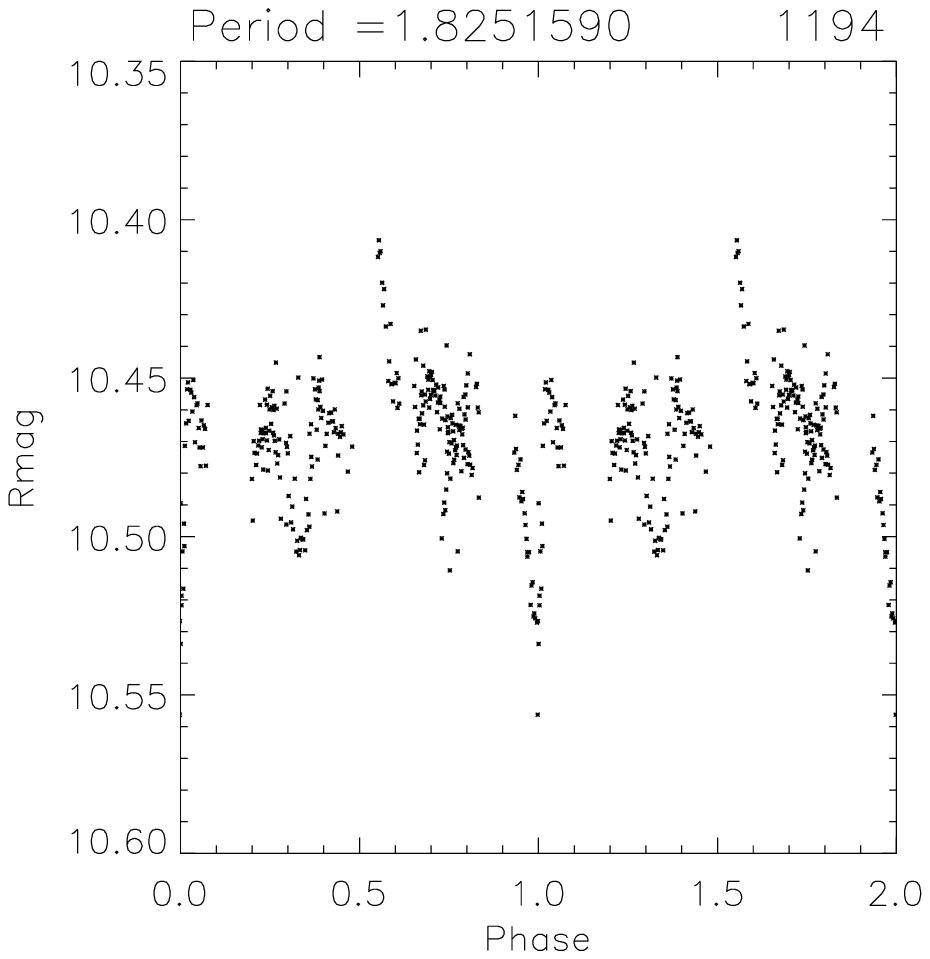}
\includegraphics[scale=.45]{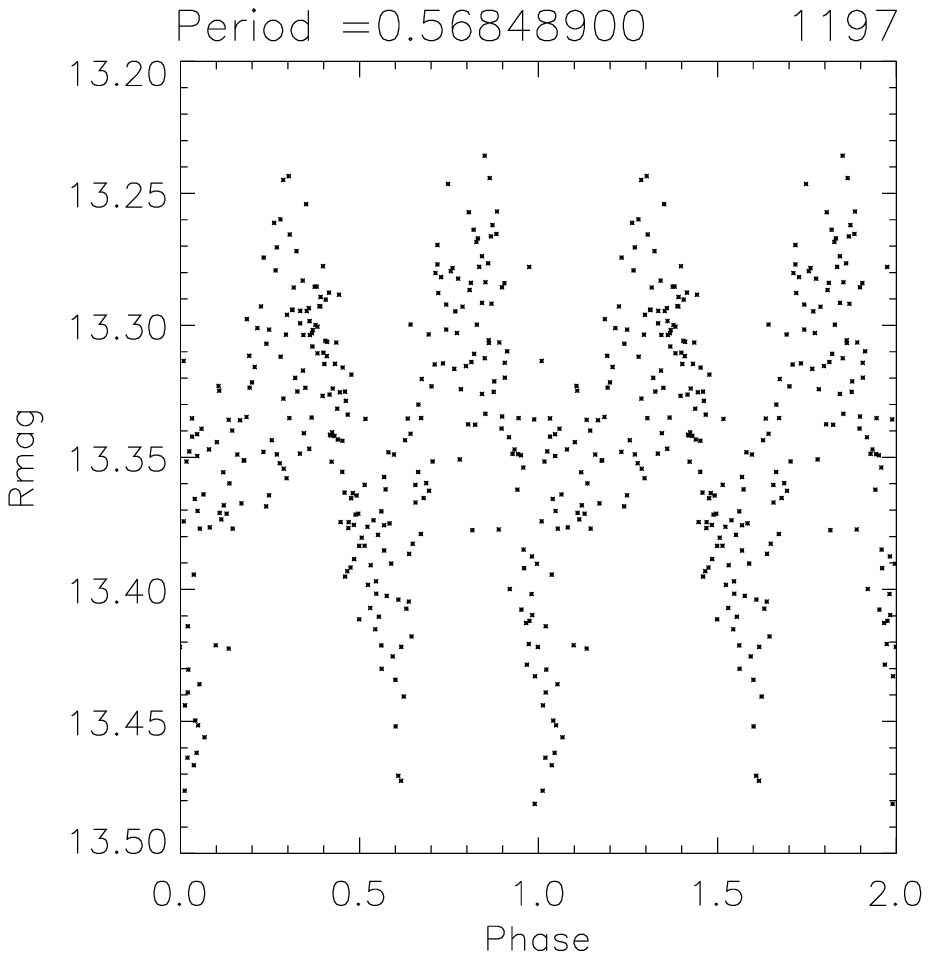}
\includegraphics[scale=.45]{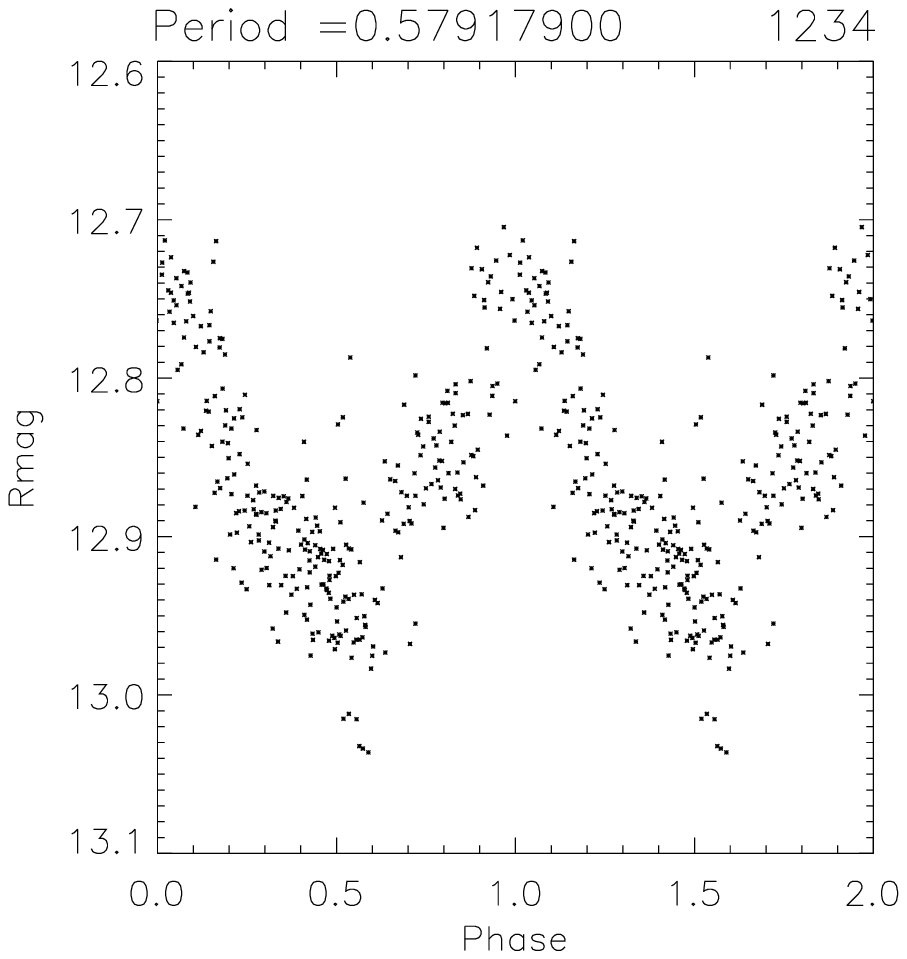}
\includegraphics[scale=.45]{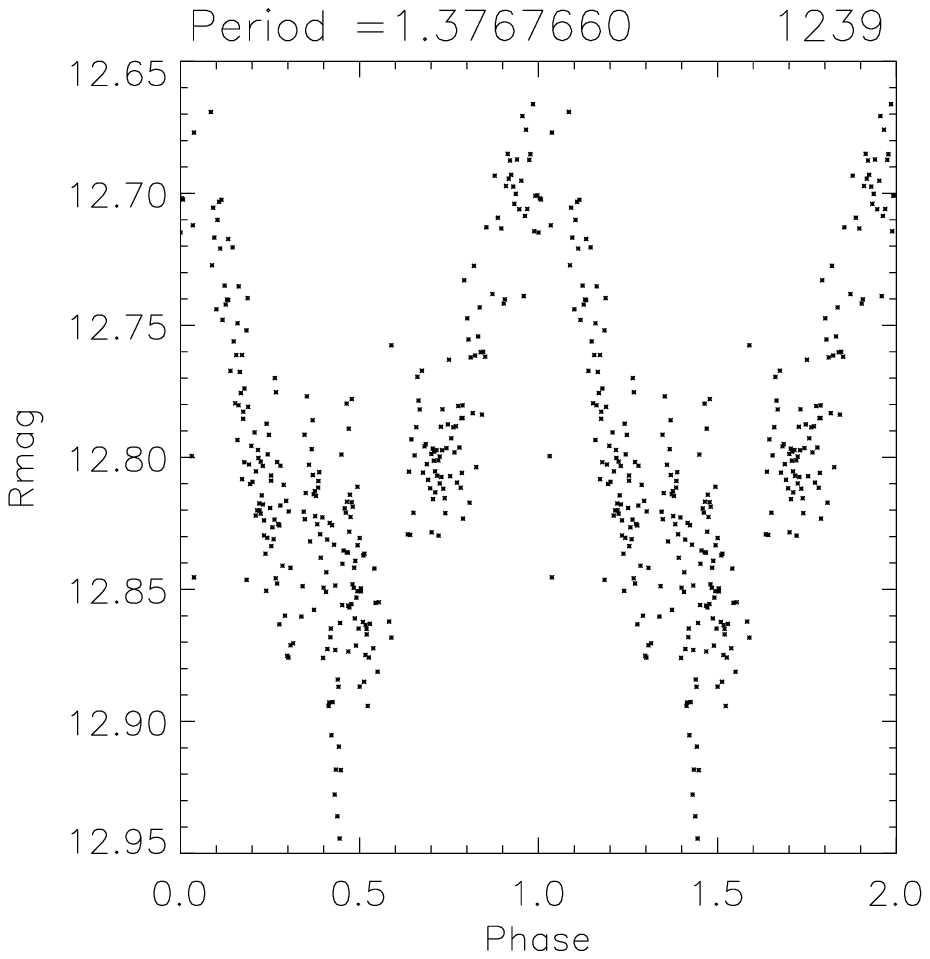}
\includegraphics[scale=.45]{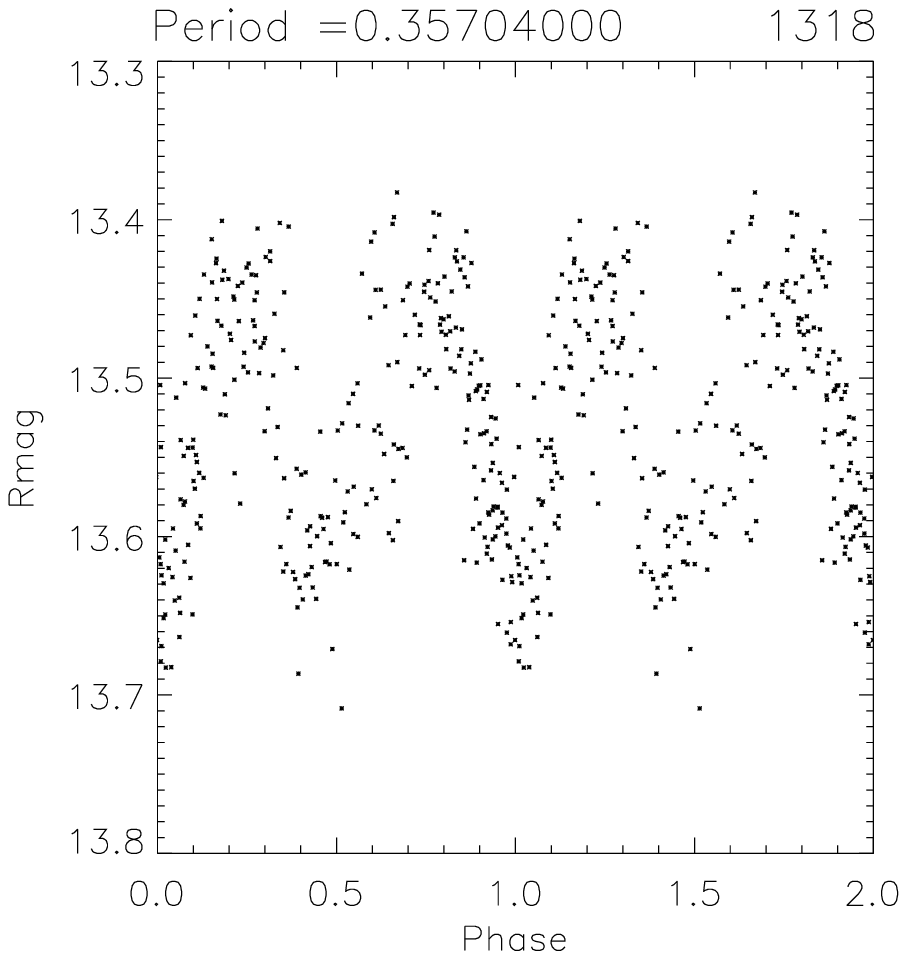}
\includegraphics[scale=.45]{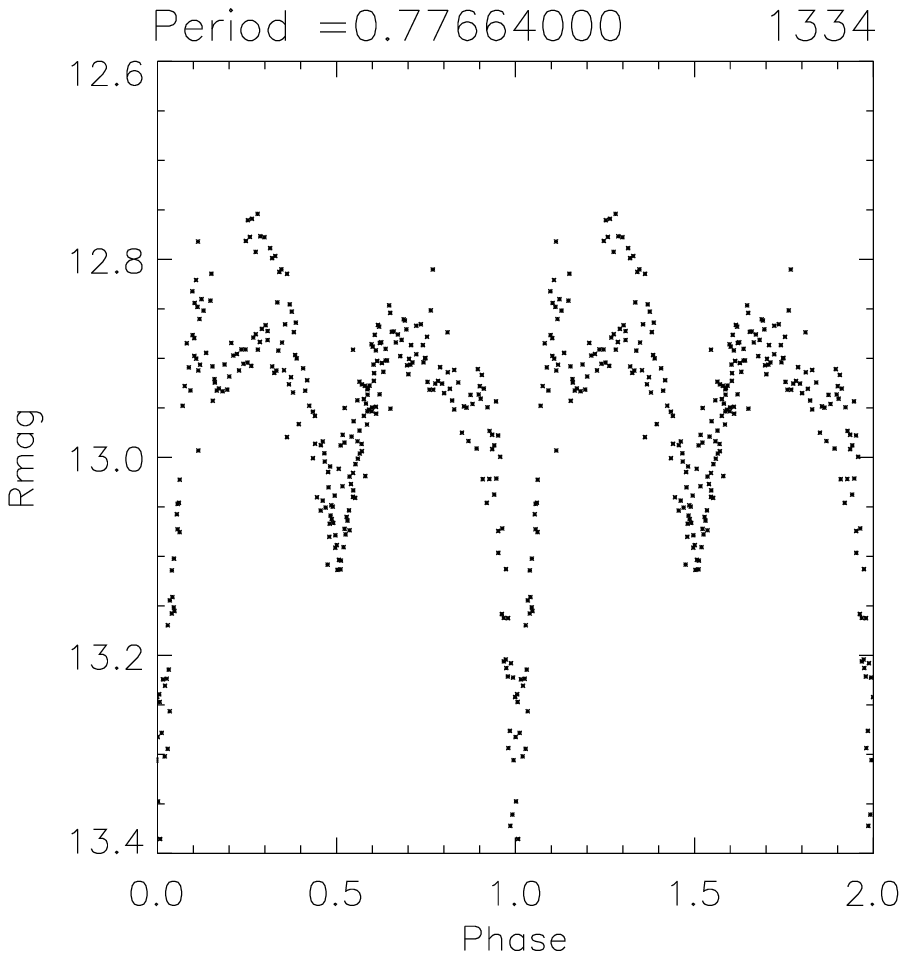}
\includegraphics[scale=.45]{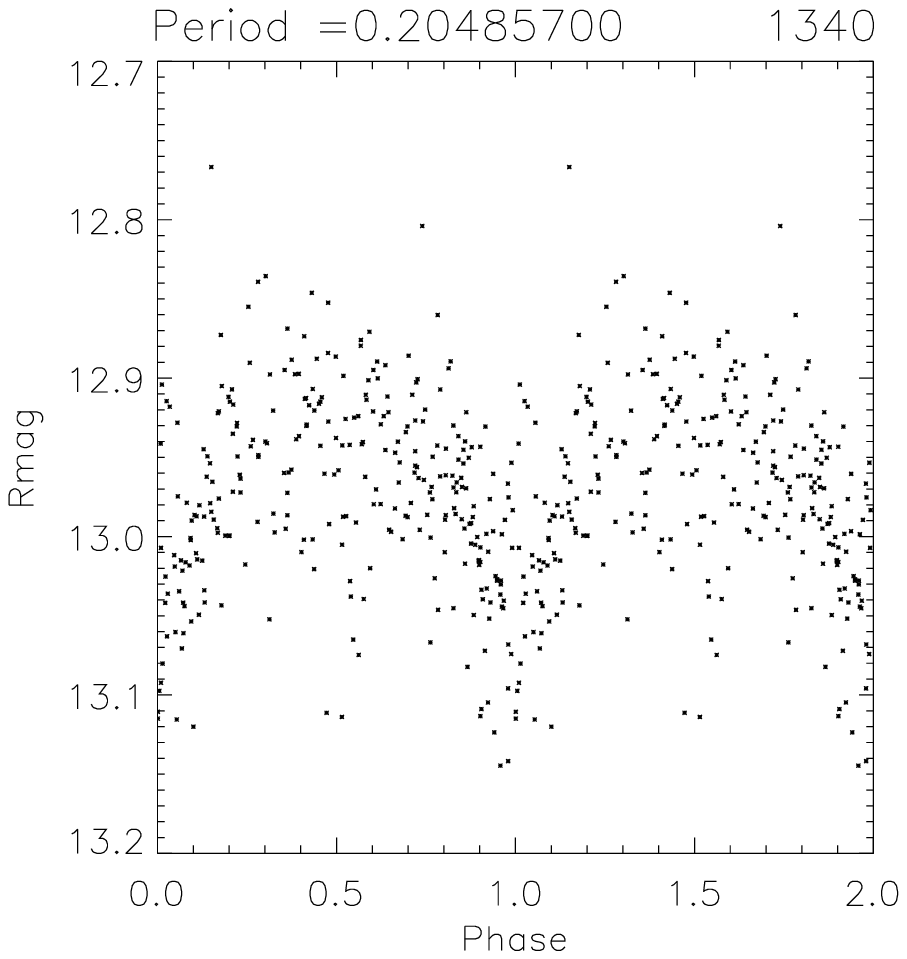}
\includegraphics[scale=.45]{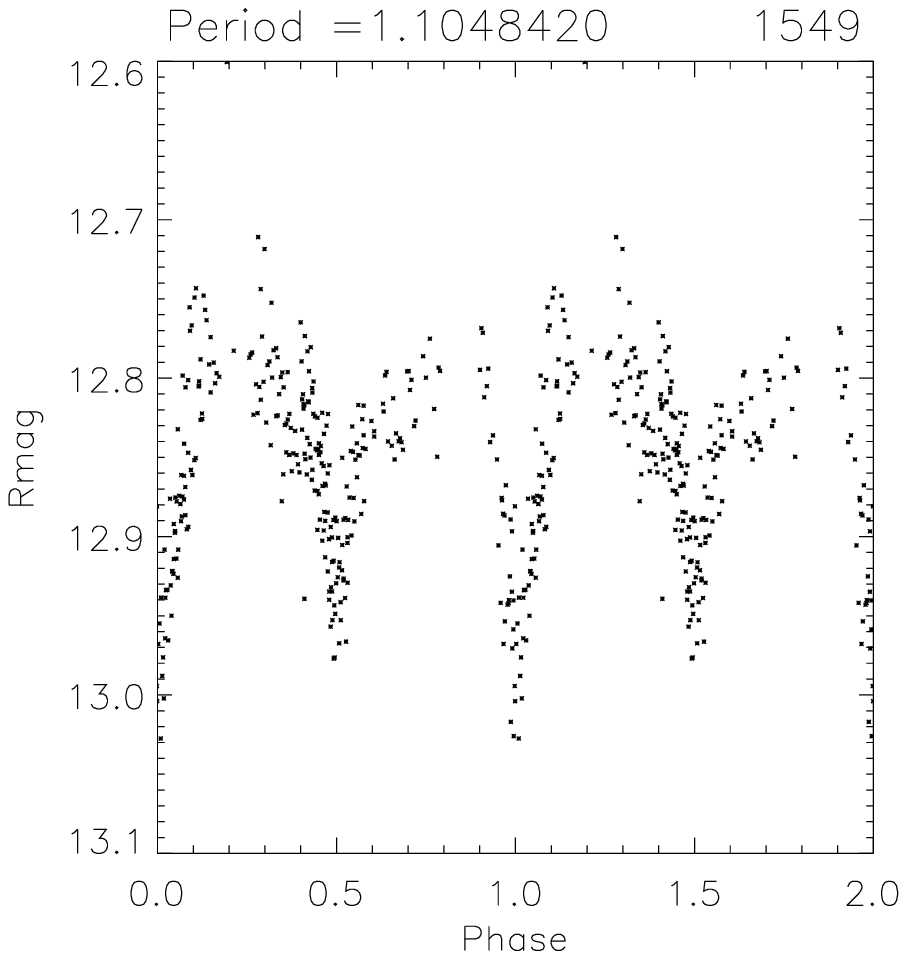}
\end{figure}
\clearpage
\begin{figure}[ht]
\includegraphics[scale=.45]{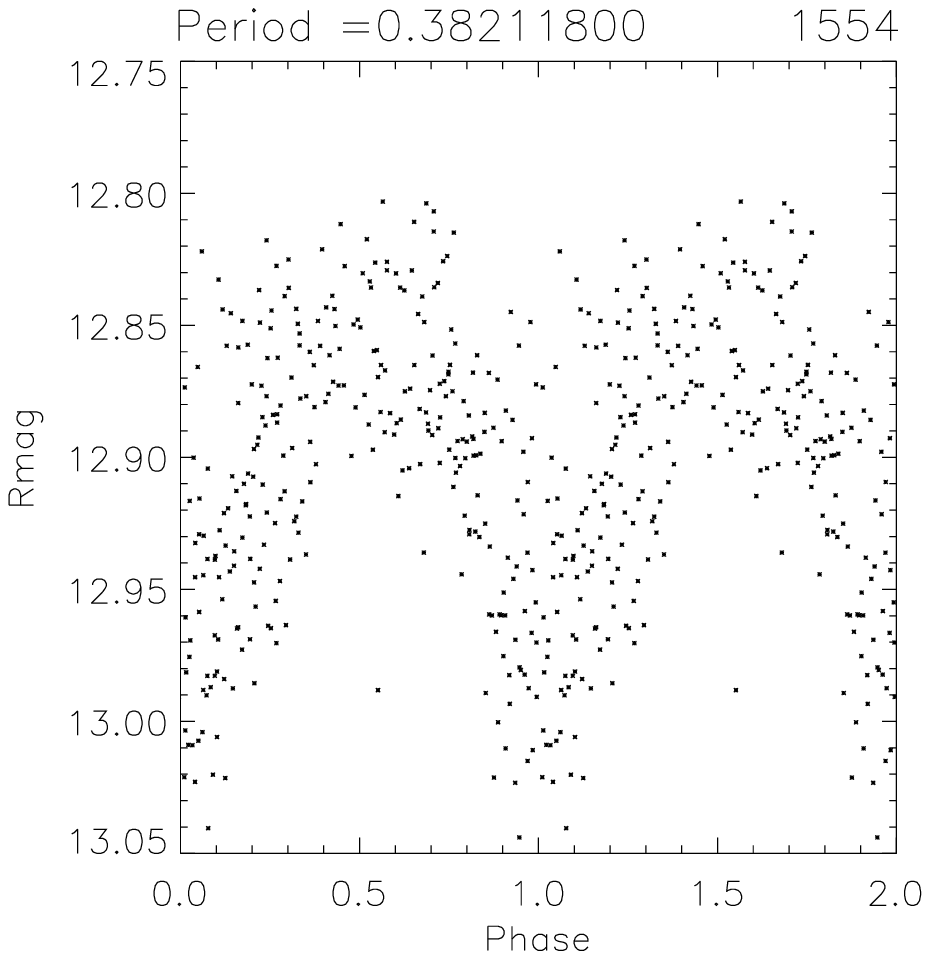}
\includegraphics[scale=.45]{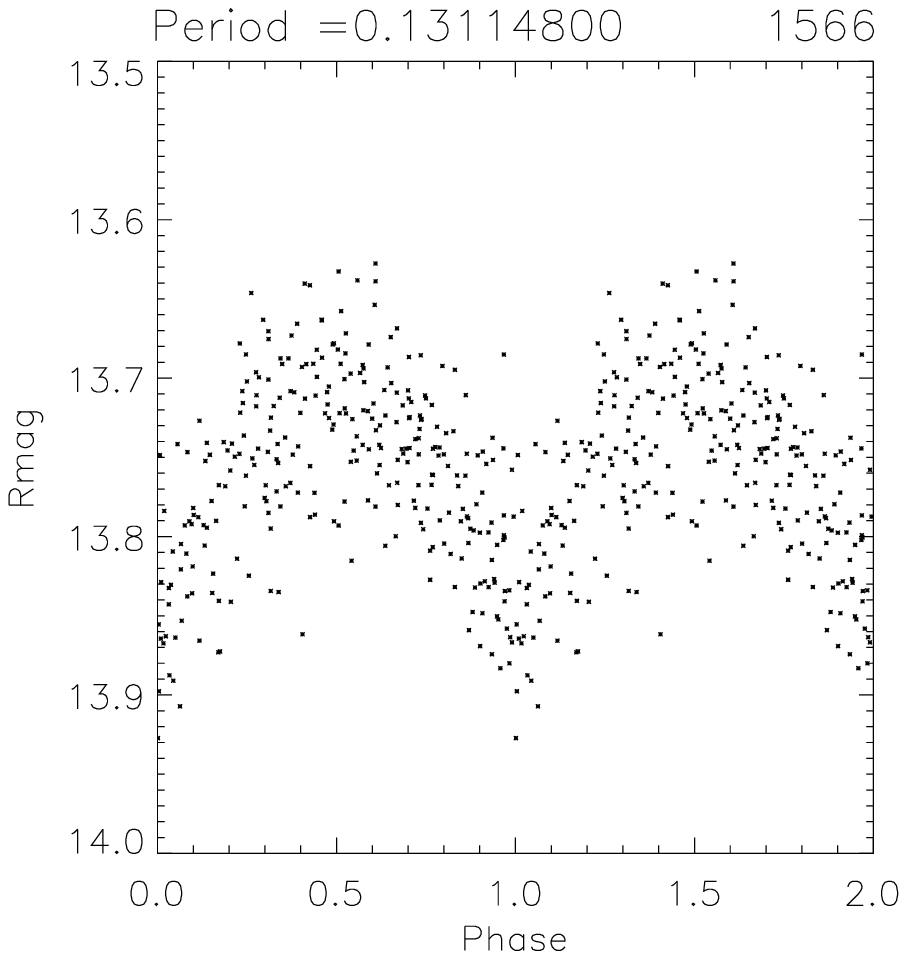}
\includegraphics[scale=.45]{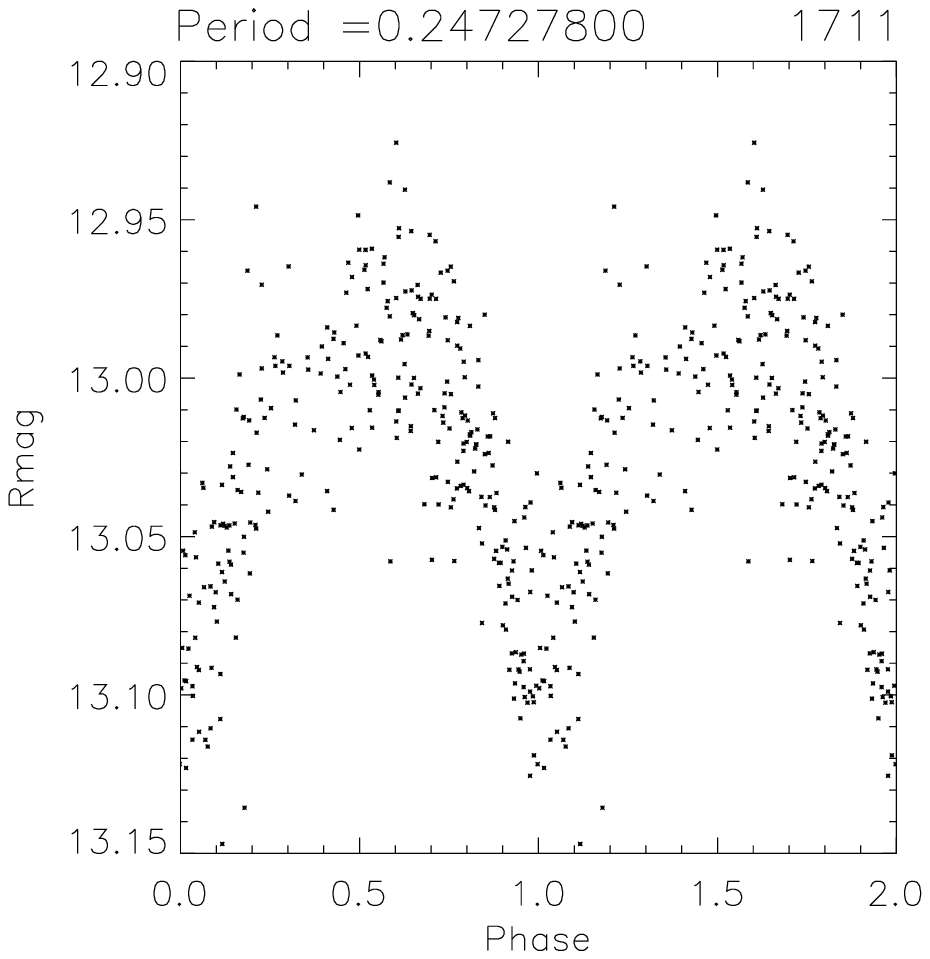}
\includegraphics[scale=.45]{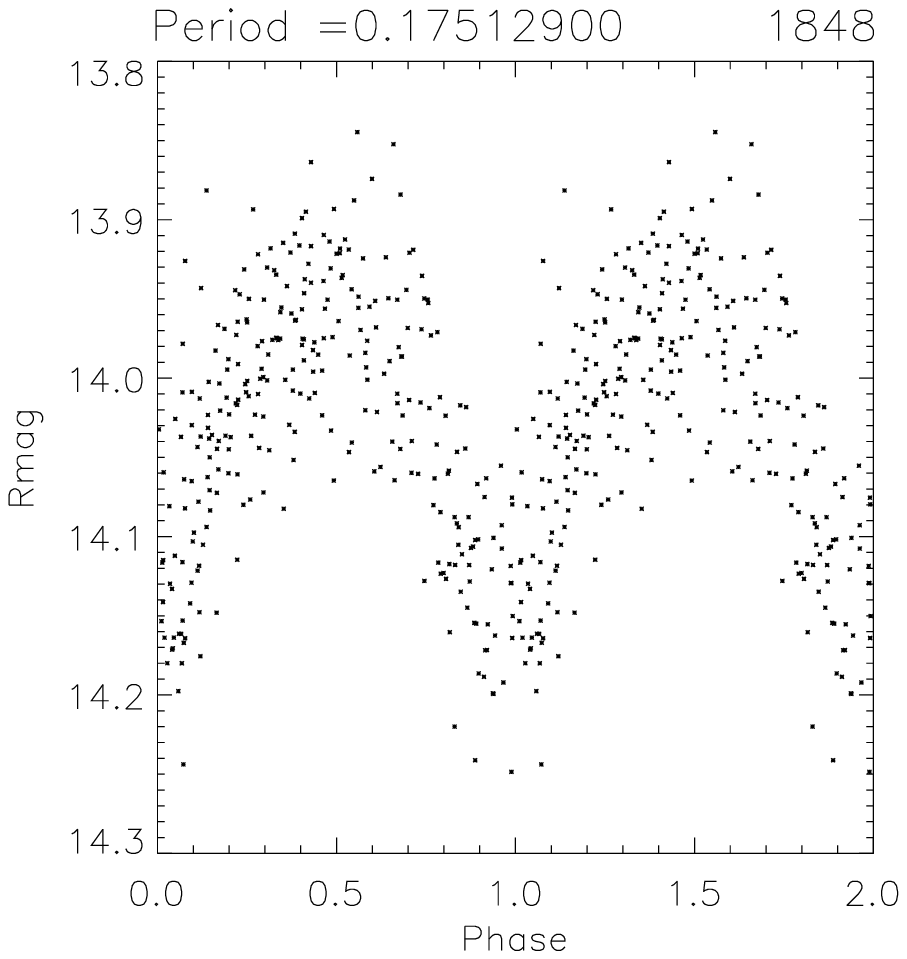}
\includegraphics[scale=.45]{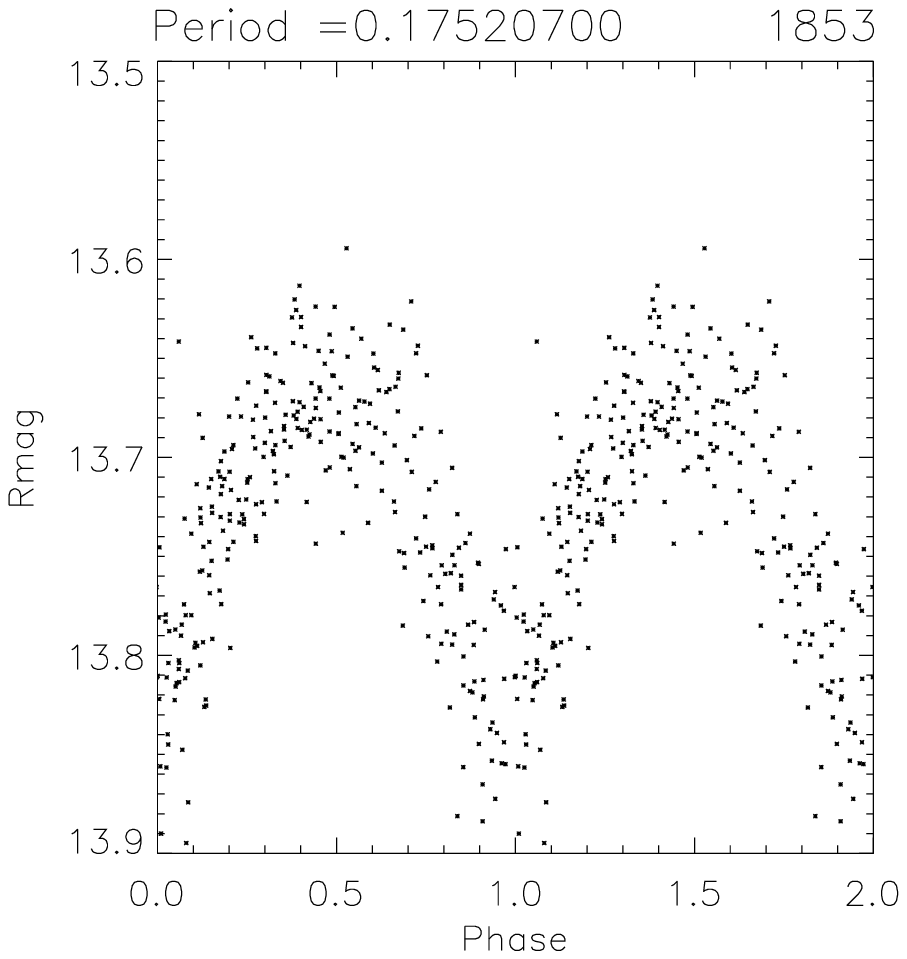}
\includegraphics[scale=.45]{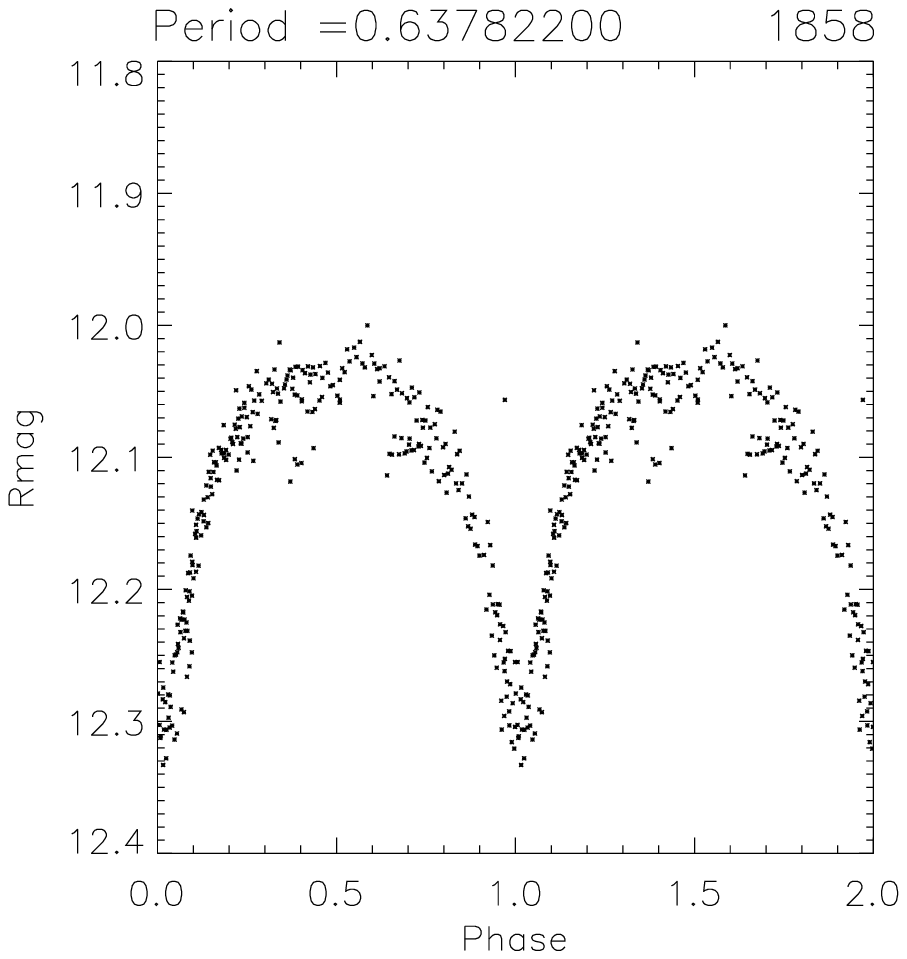}
\includegraphics[scale=.45]{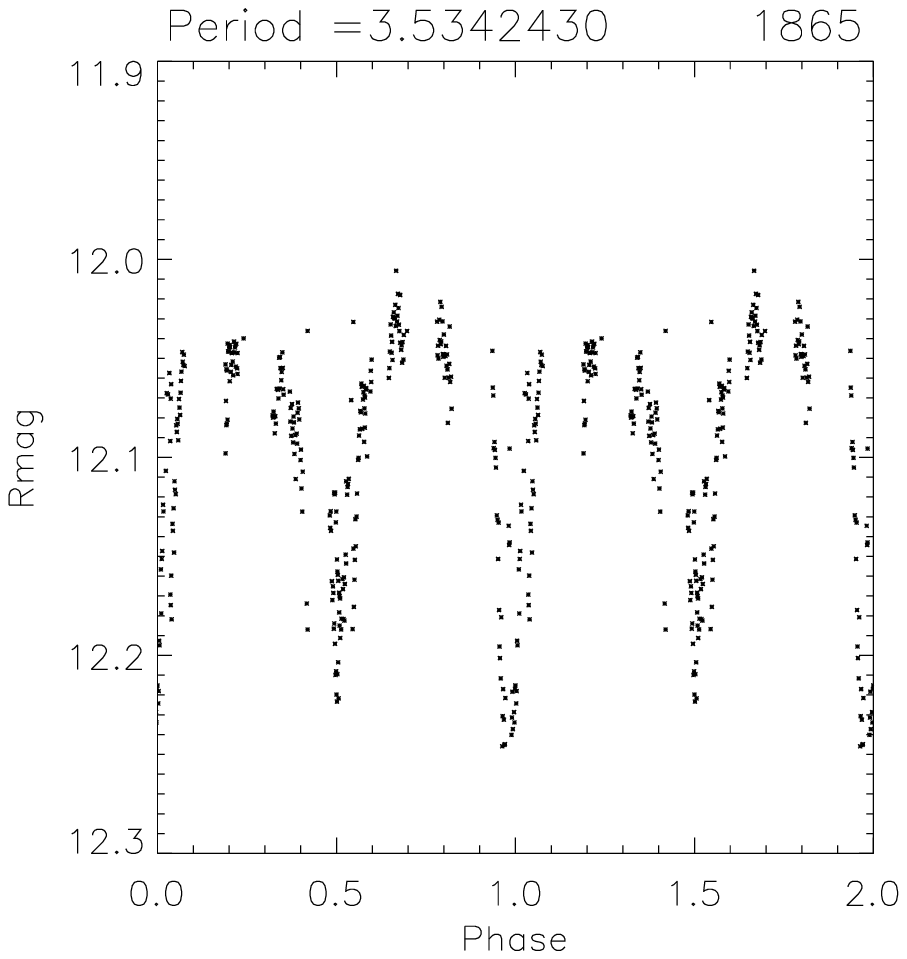}
\includegraphics[scale=.45]{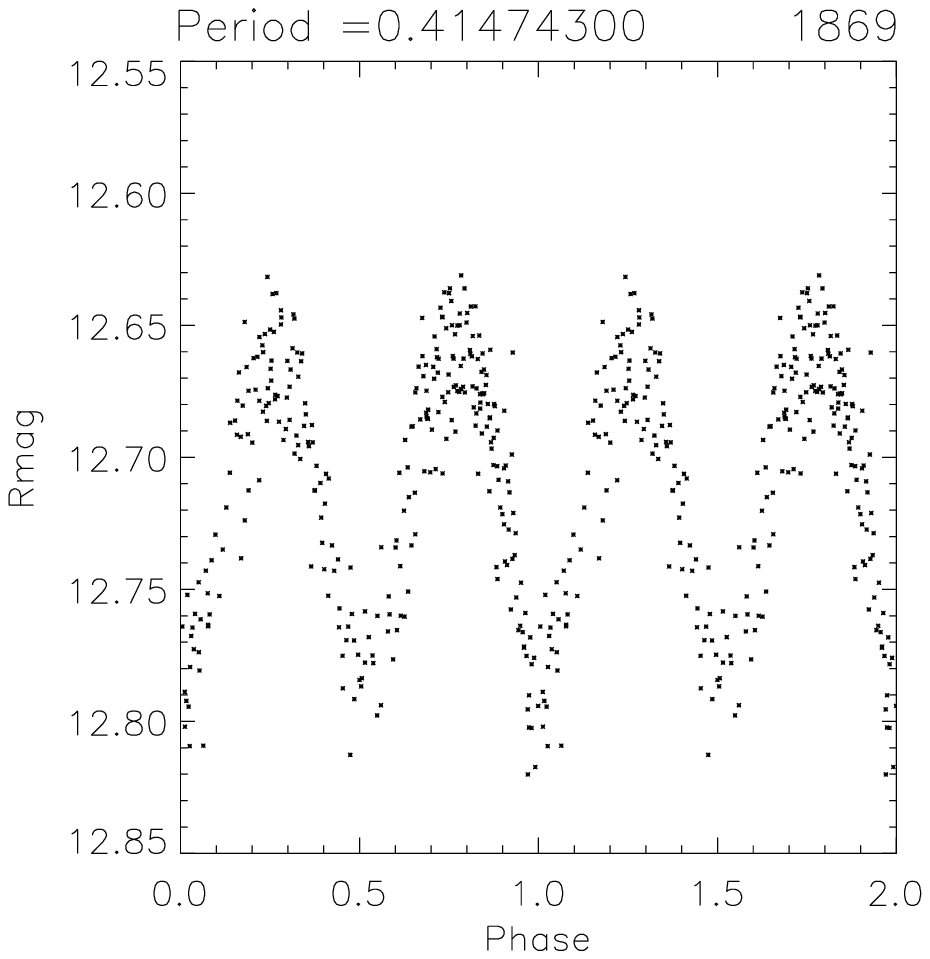}
\includegraphics[scale=.45]{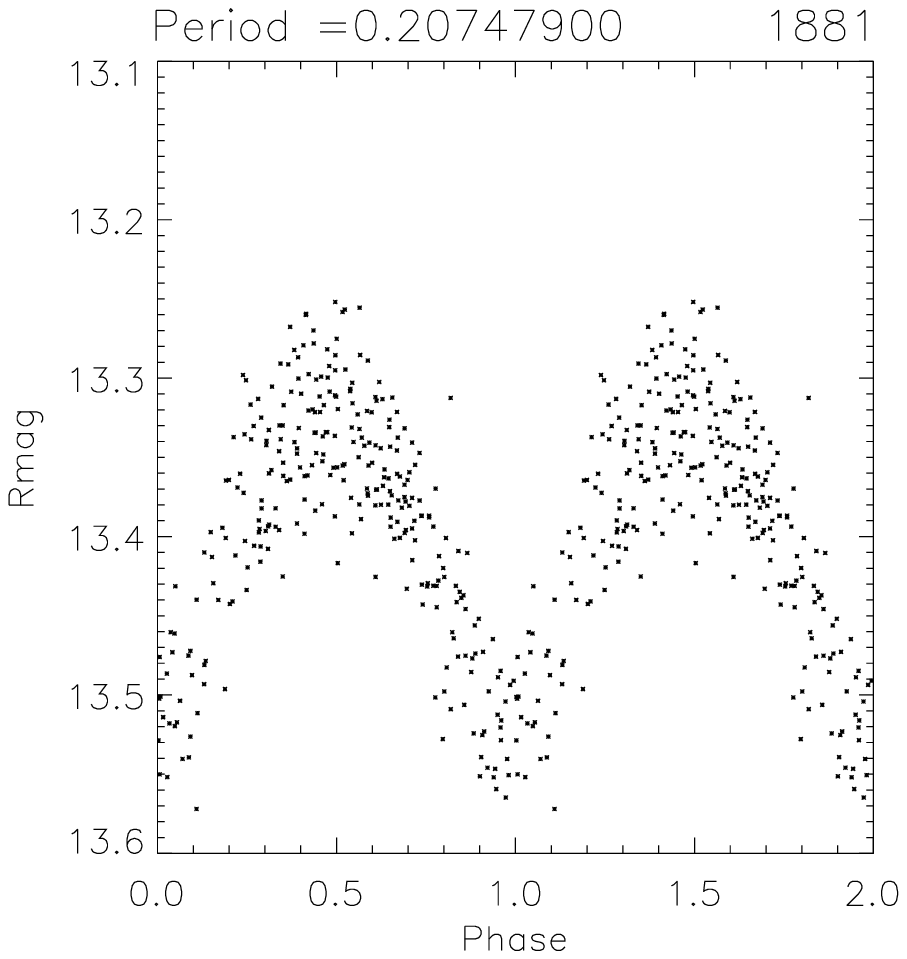}
\includegraphics[scale=.45]{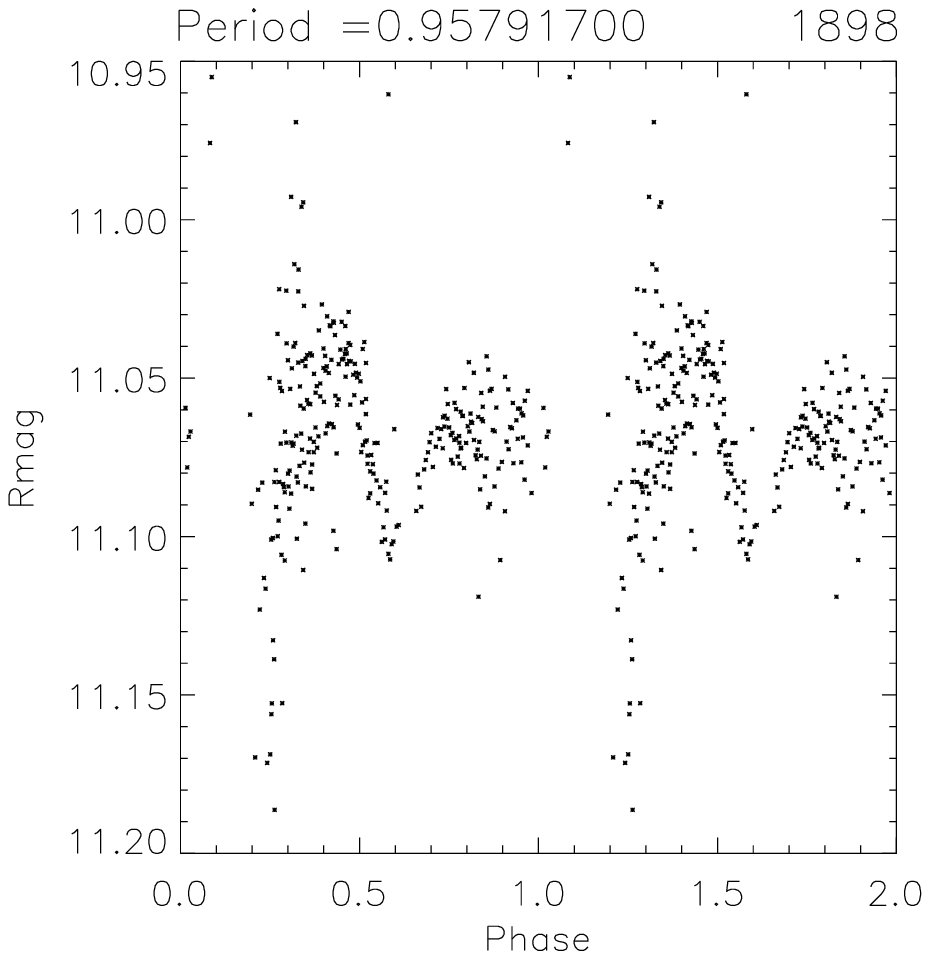}
\includegraphics[scale=.45]{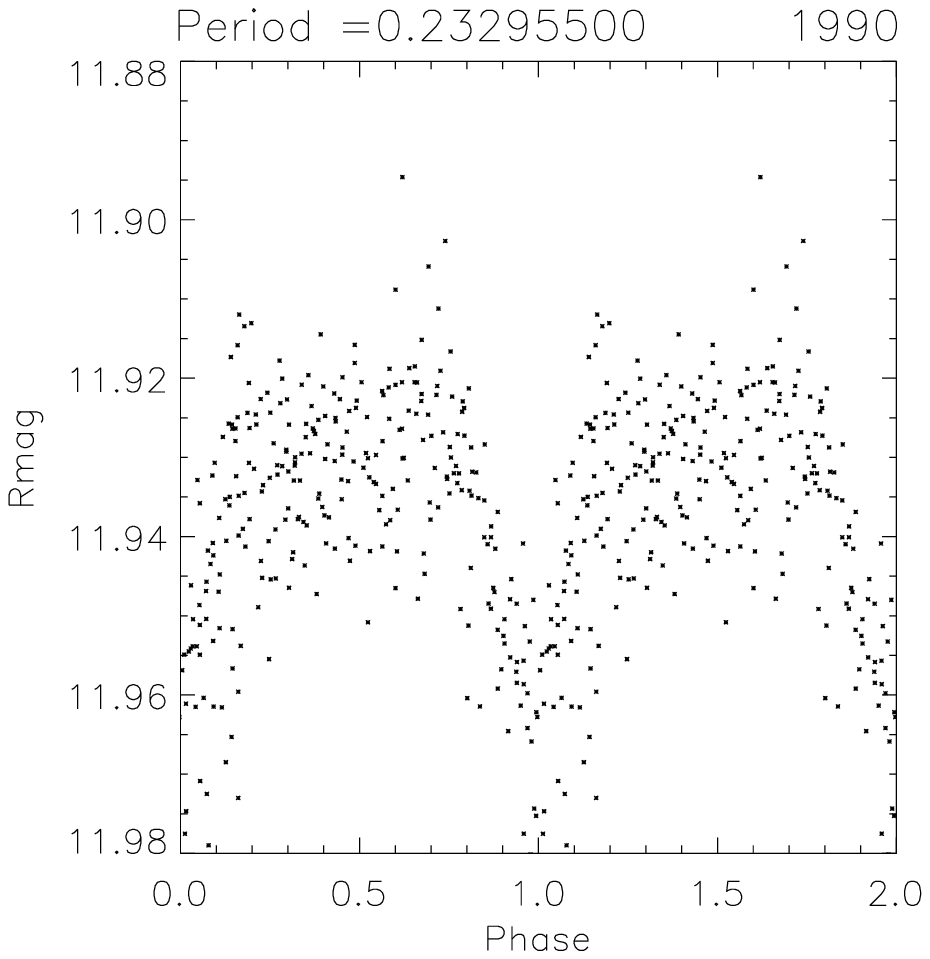}
\includegraphics[scale=.45]{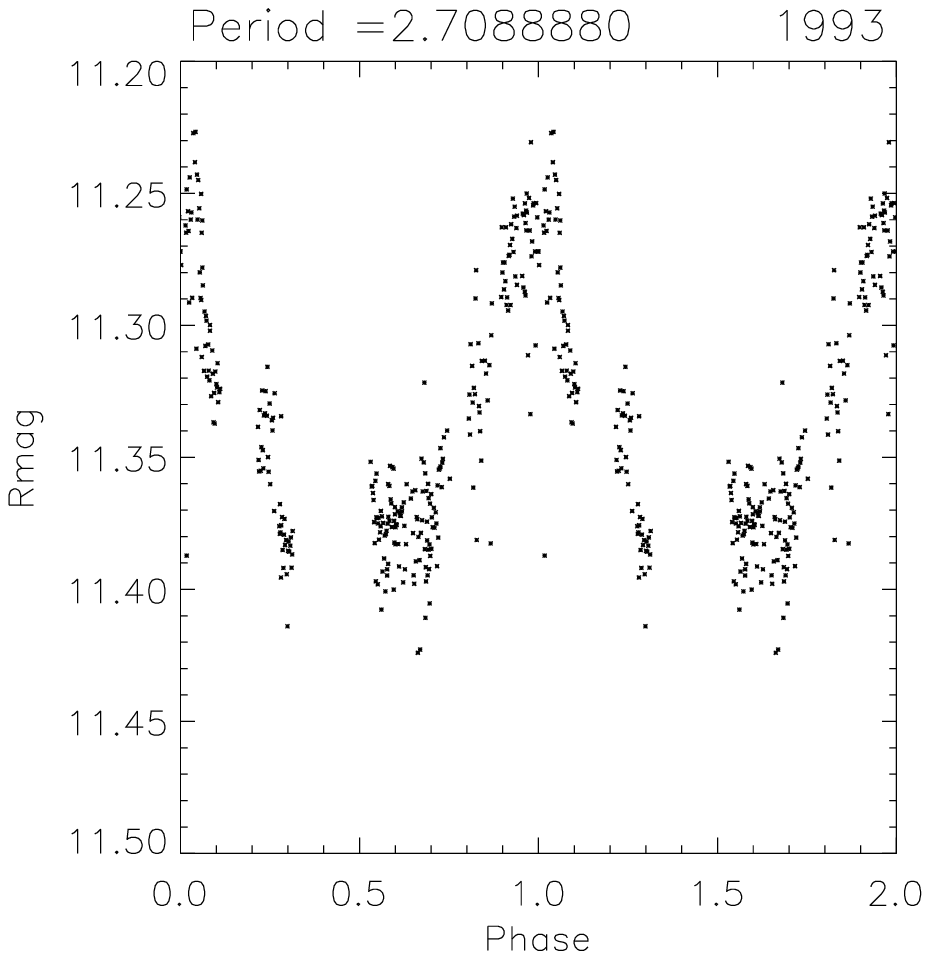}
\end{figure}
\clearpage
\begin{figure}[ht]
\includegraphics[scale=.45]{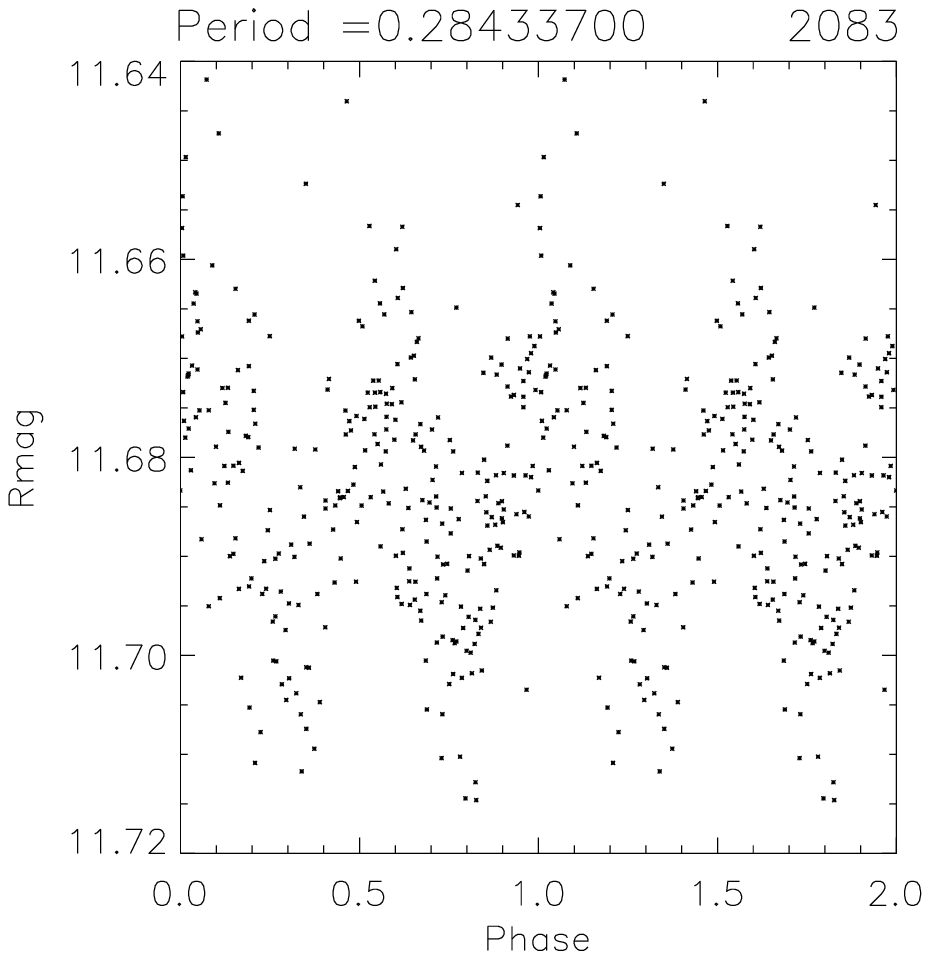}
\includegraphics[scale=.45]{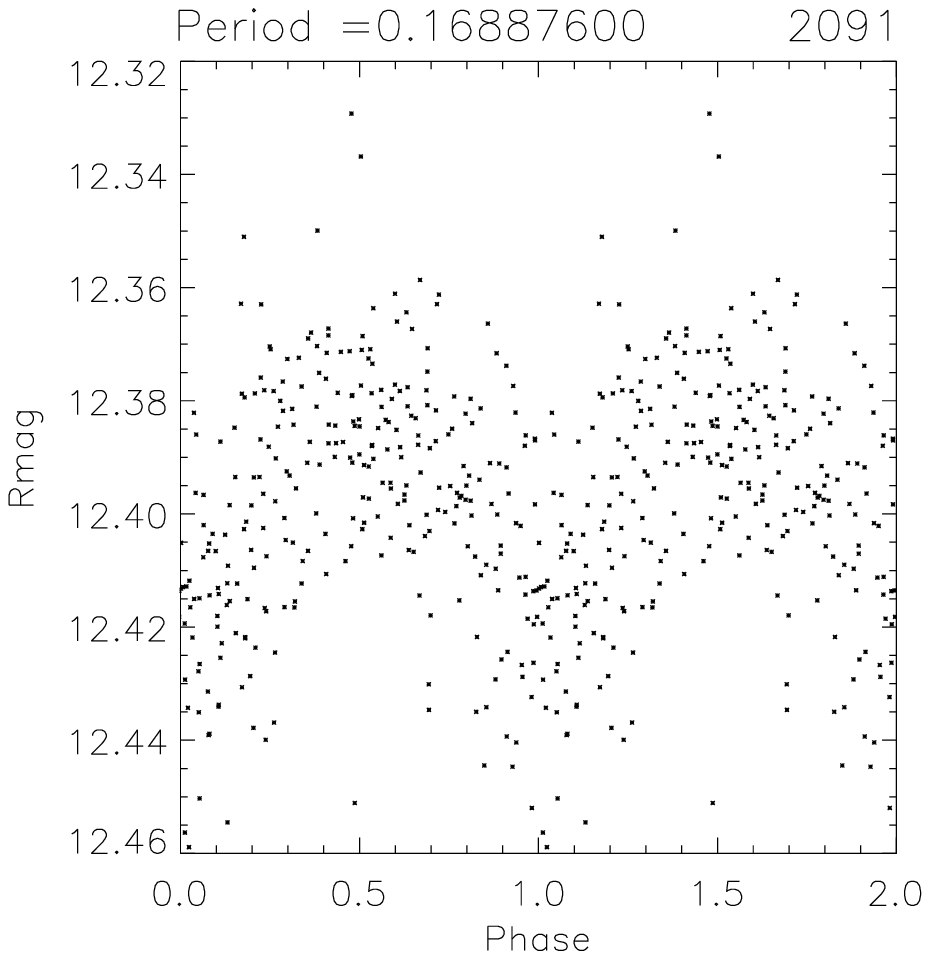}
\includegraphics[scale=.45]{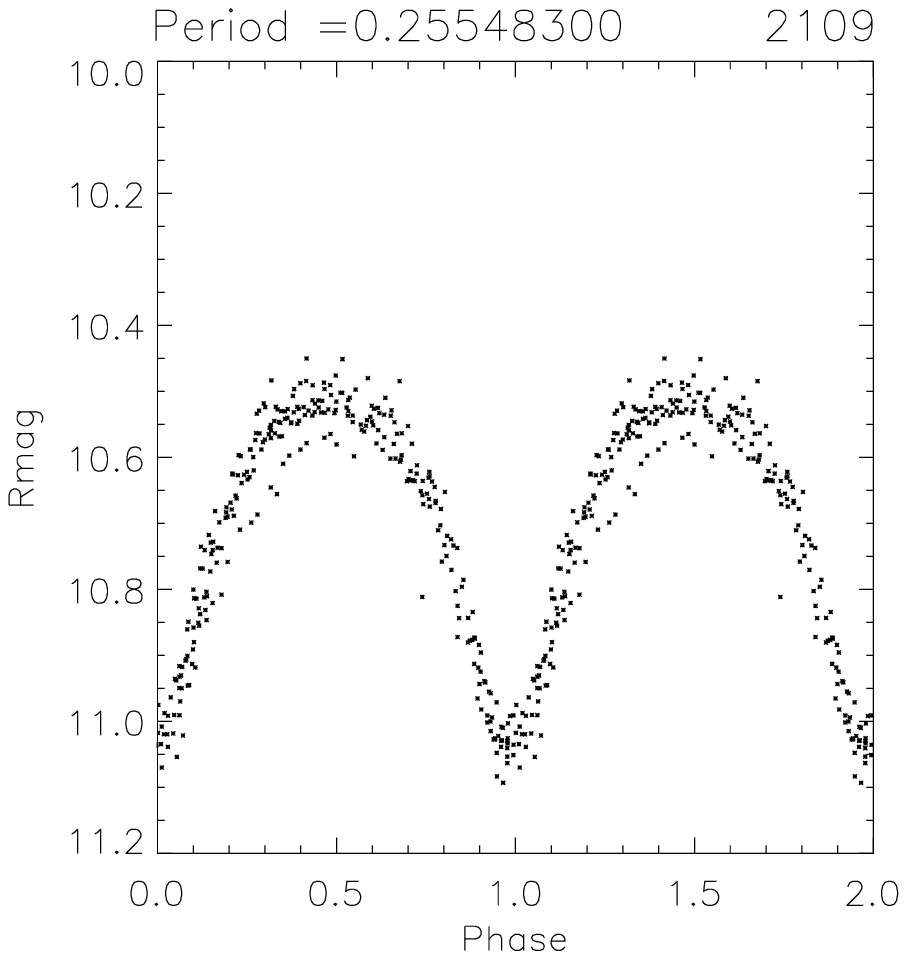}
\includegraphics[scale=.45]{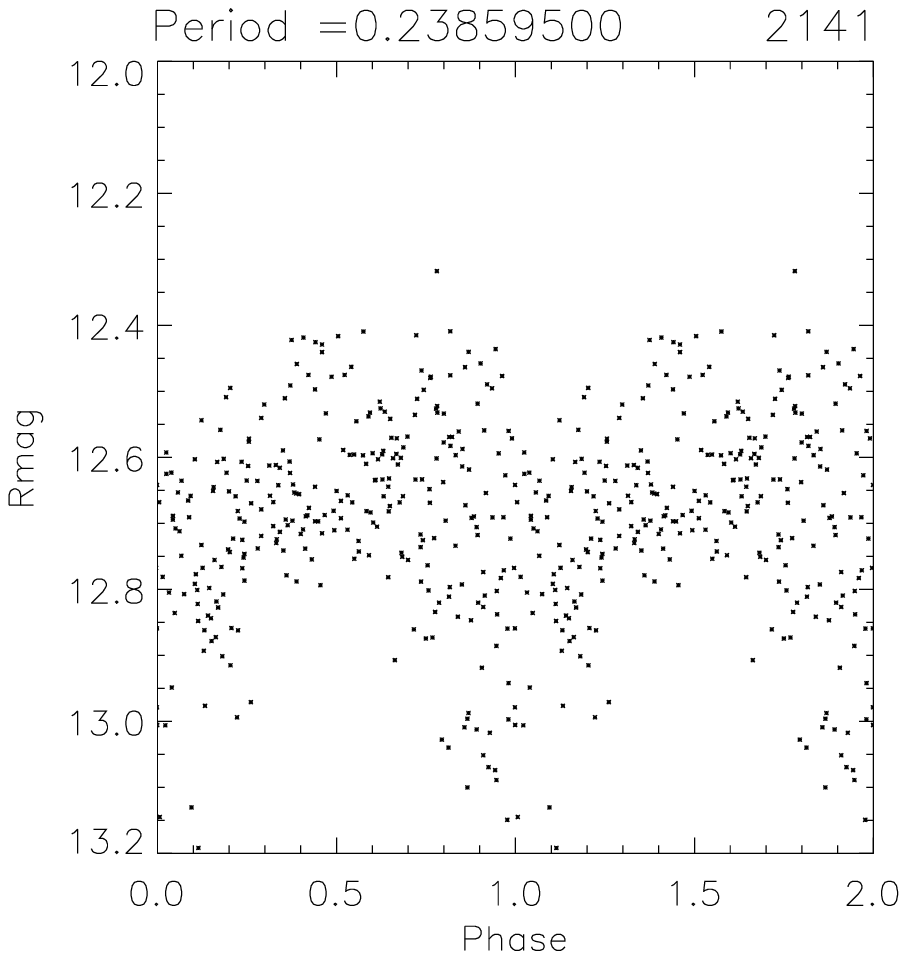}
\includegraphics[scale=.45]{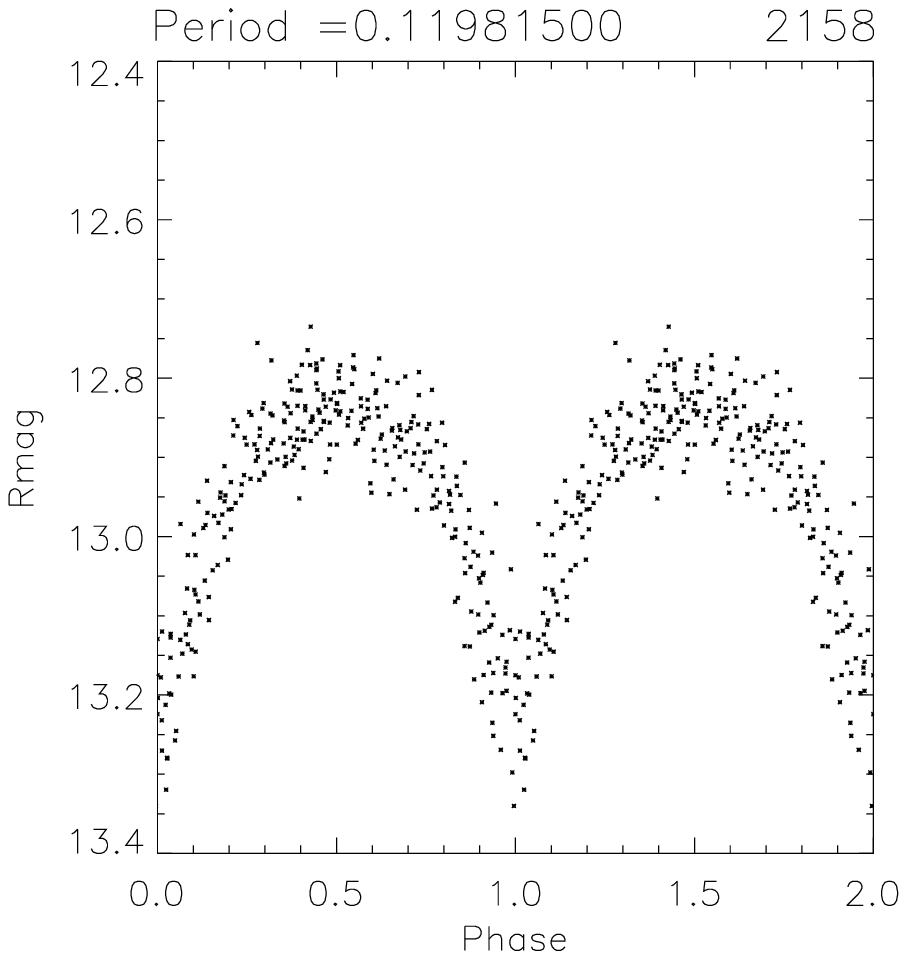}
\includegraphics[scale=.45]{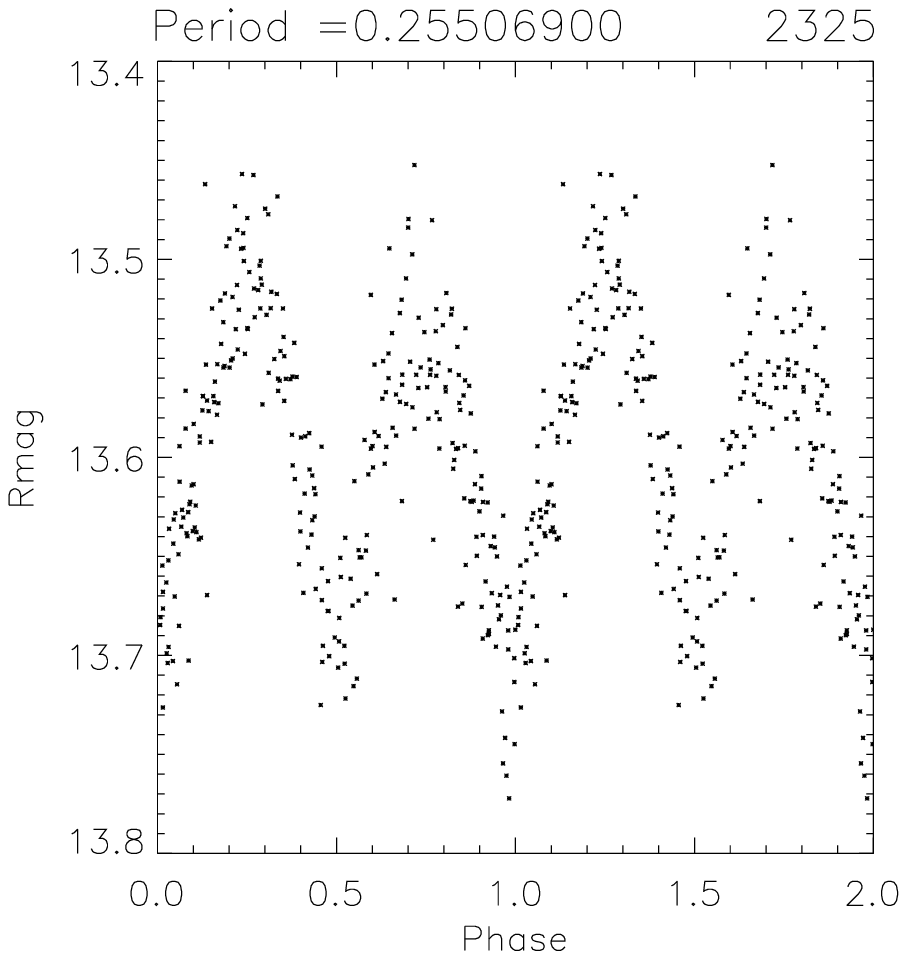}
\includegraphics[scale=.45]{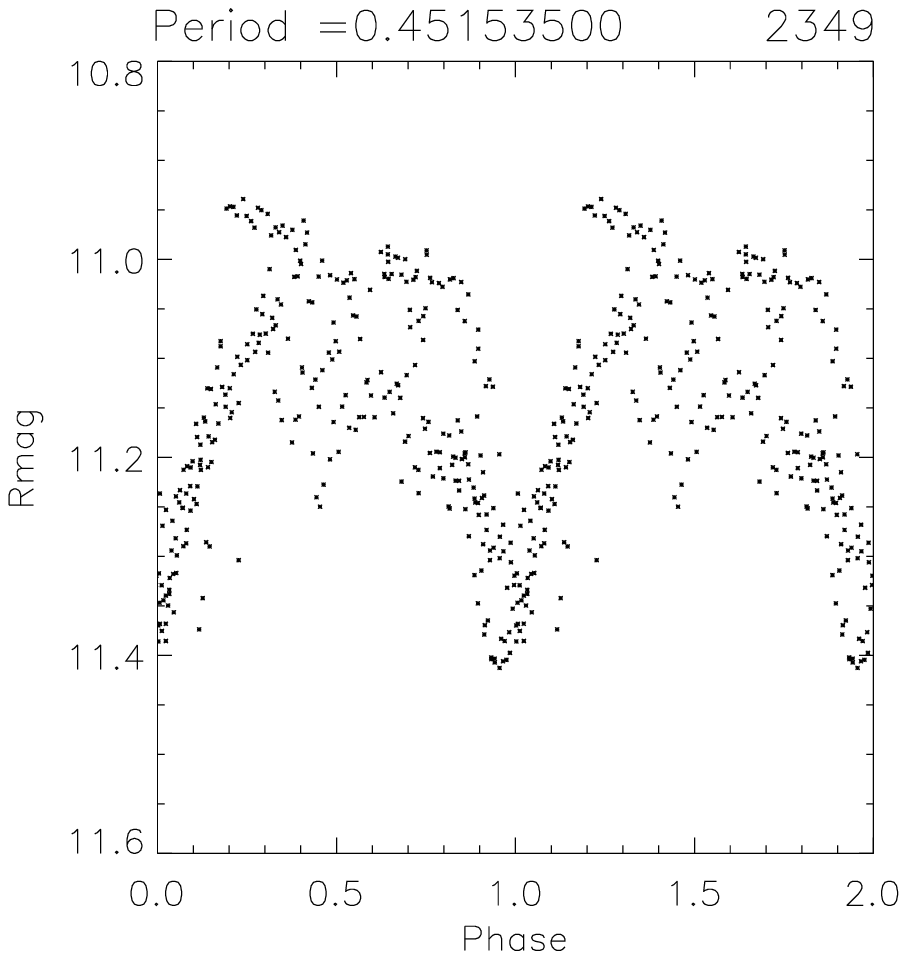}
\includegraphics[scale=.45]{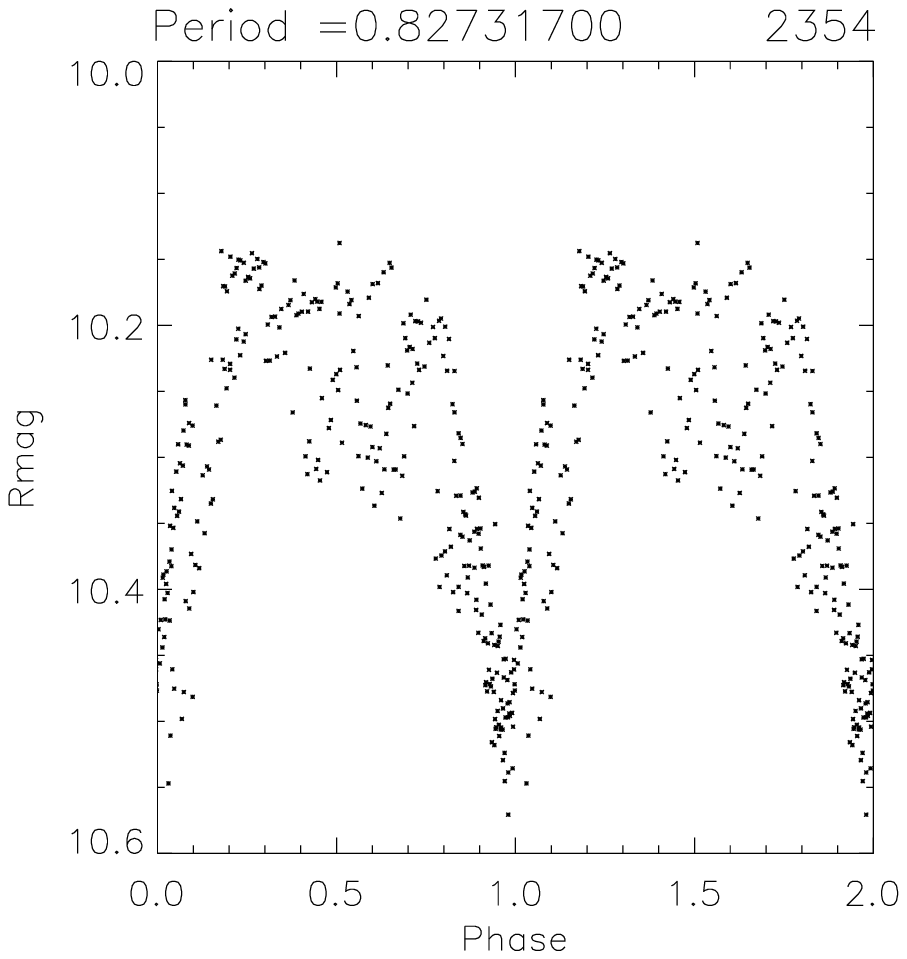}
\includegraphics[scale=.45]{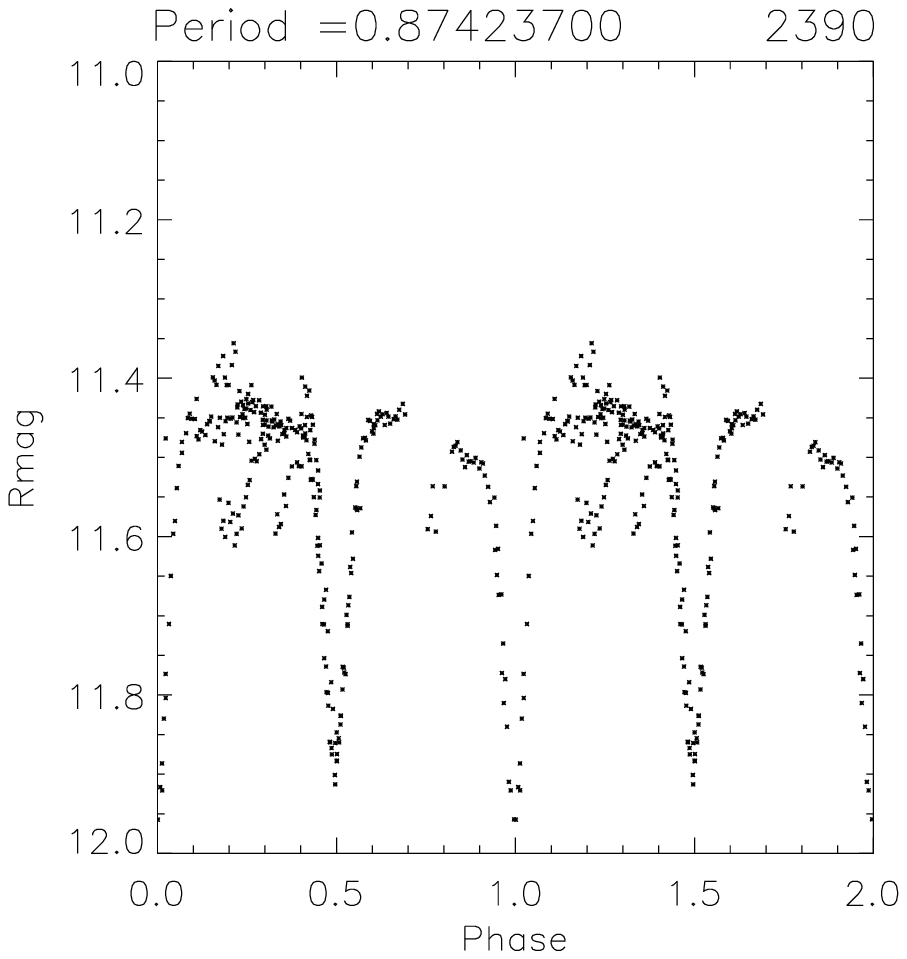}
\includegraphics[scale=.45]{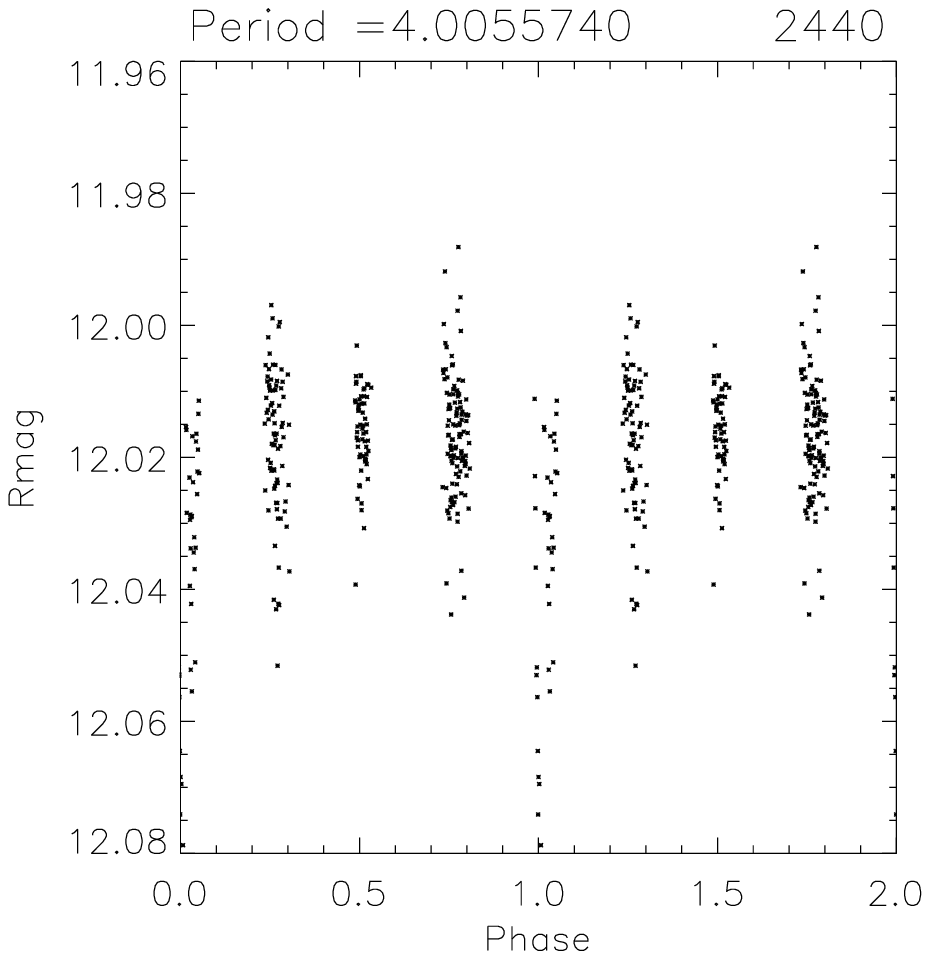}
\includegraphics[scale=.45]{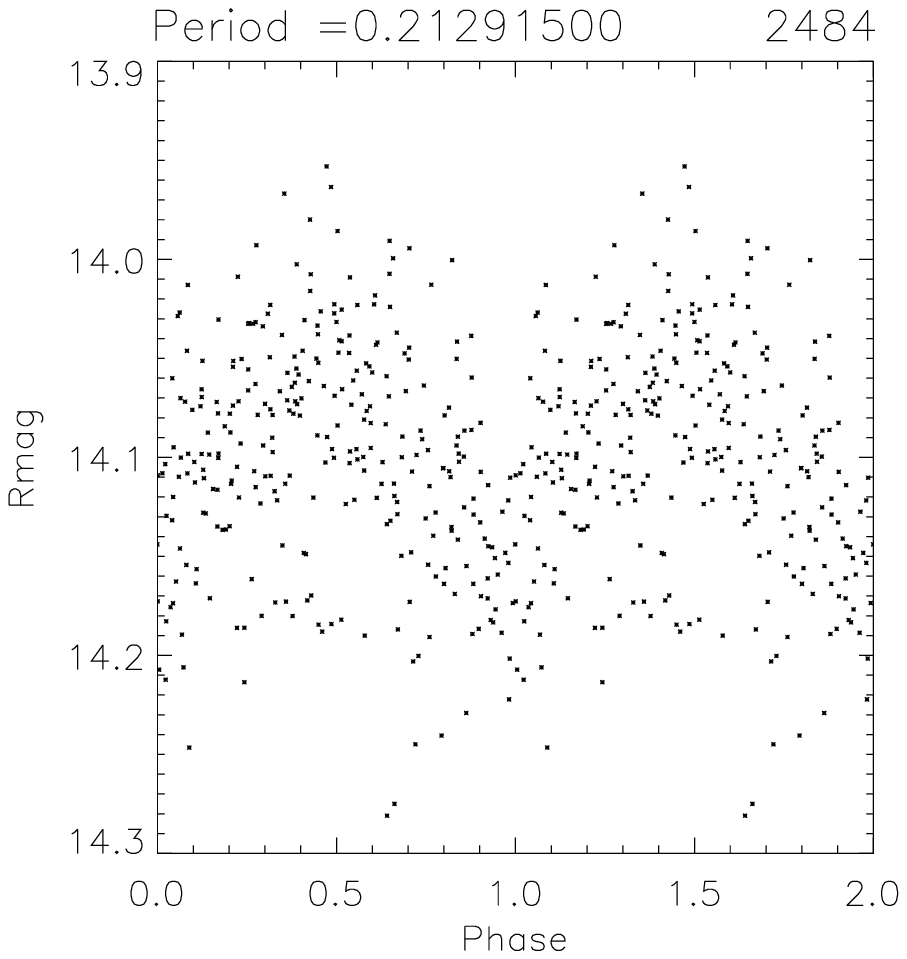}
\includegraphics[scale=.45]{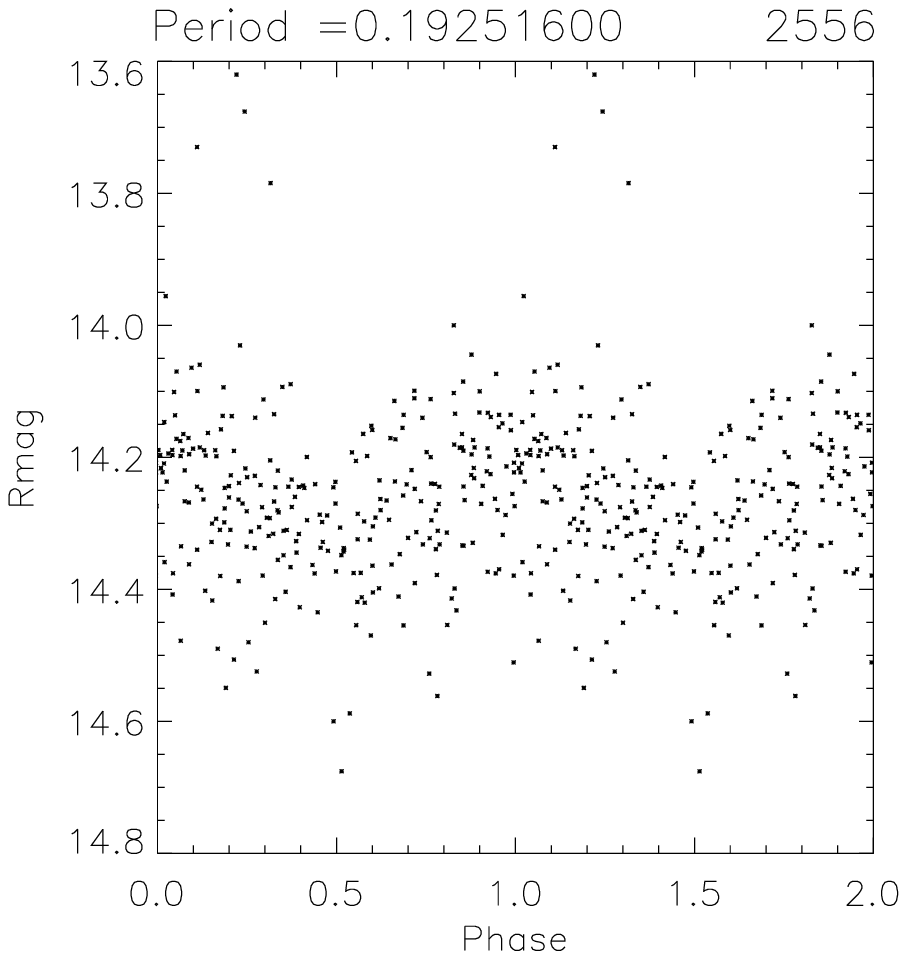}
\end{figure}
\clearpage
\begin{figure}[ht]
\includegraphics[scale=.45]{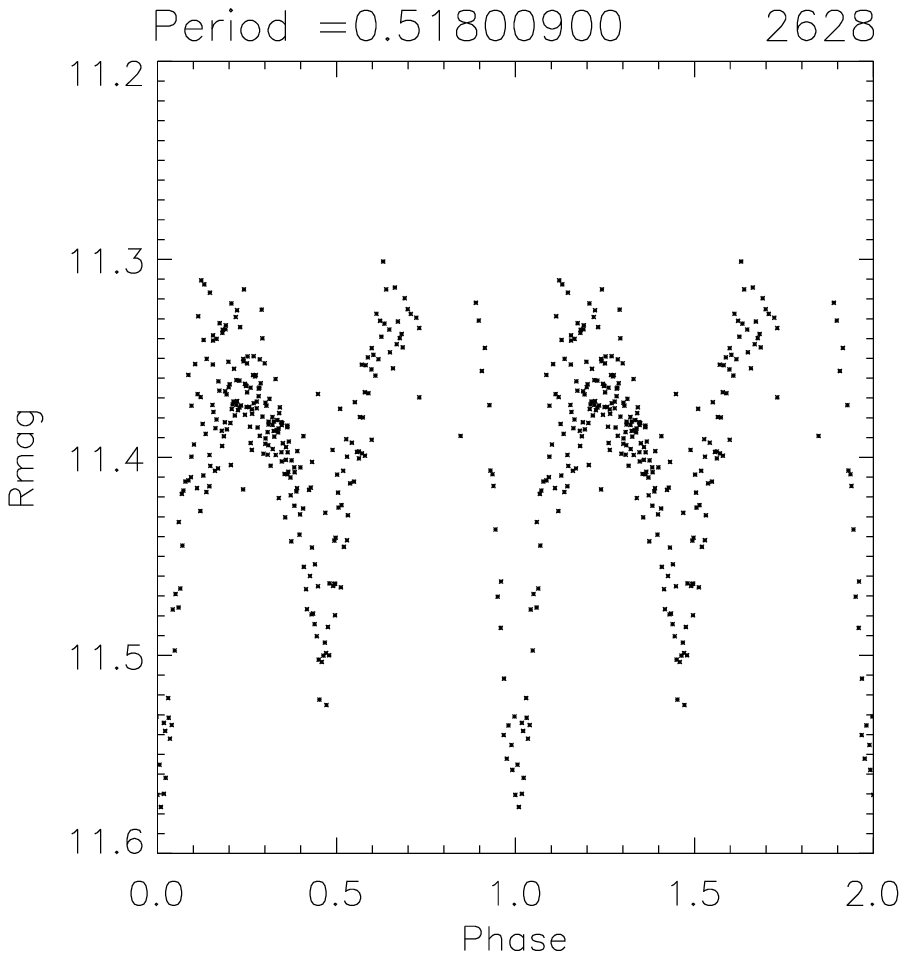}
\includegraphics[scale=.45]{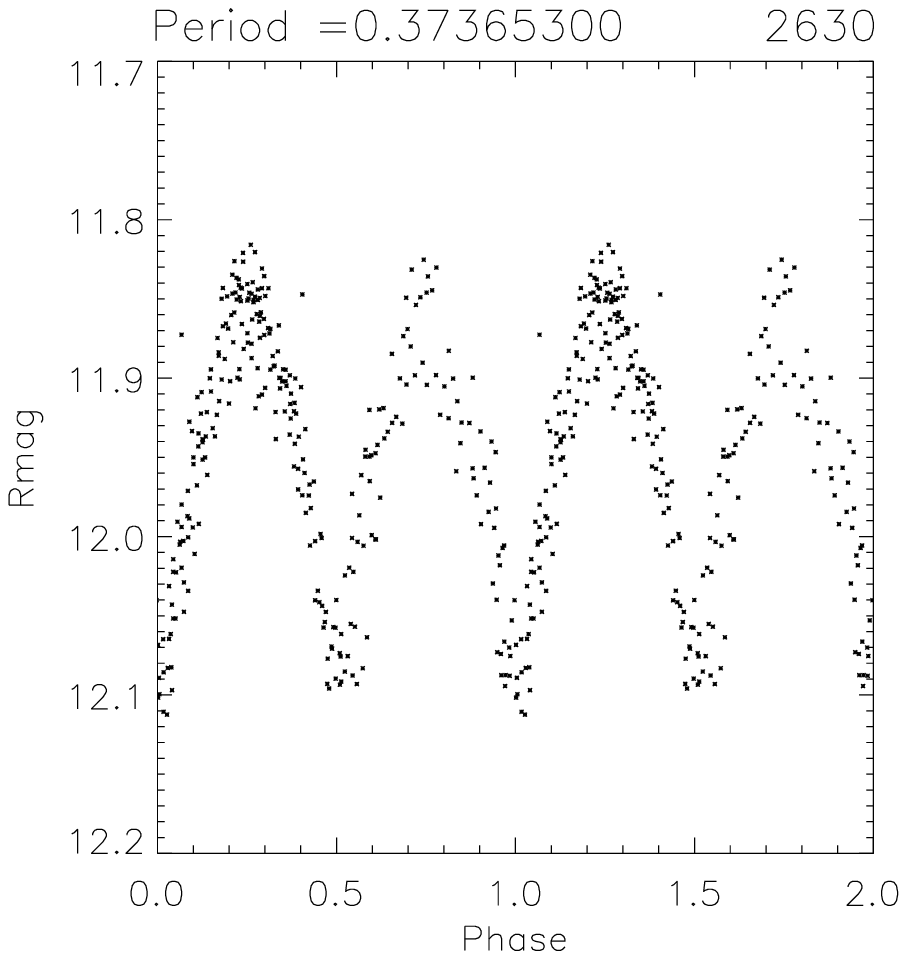}
\includegraphics[scale=.45]{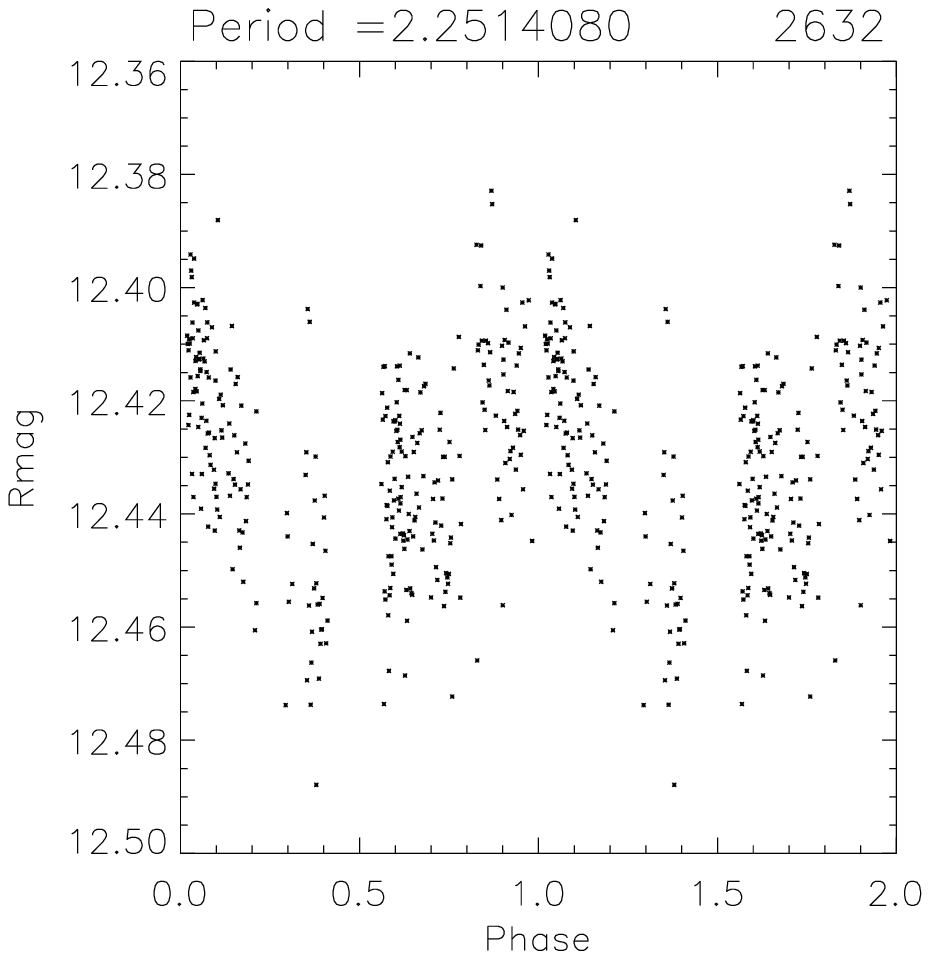}
\includegraphics[scale=.45]{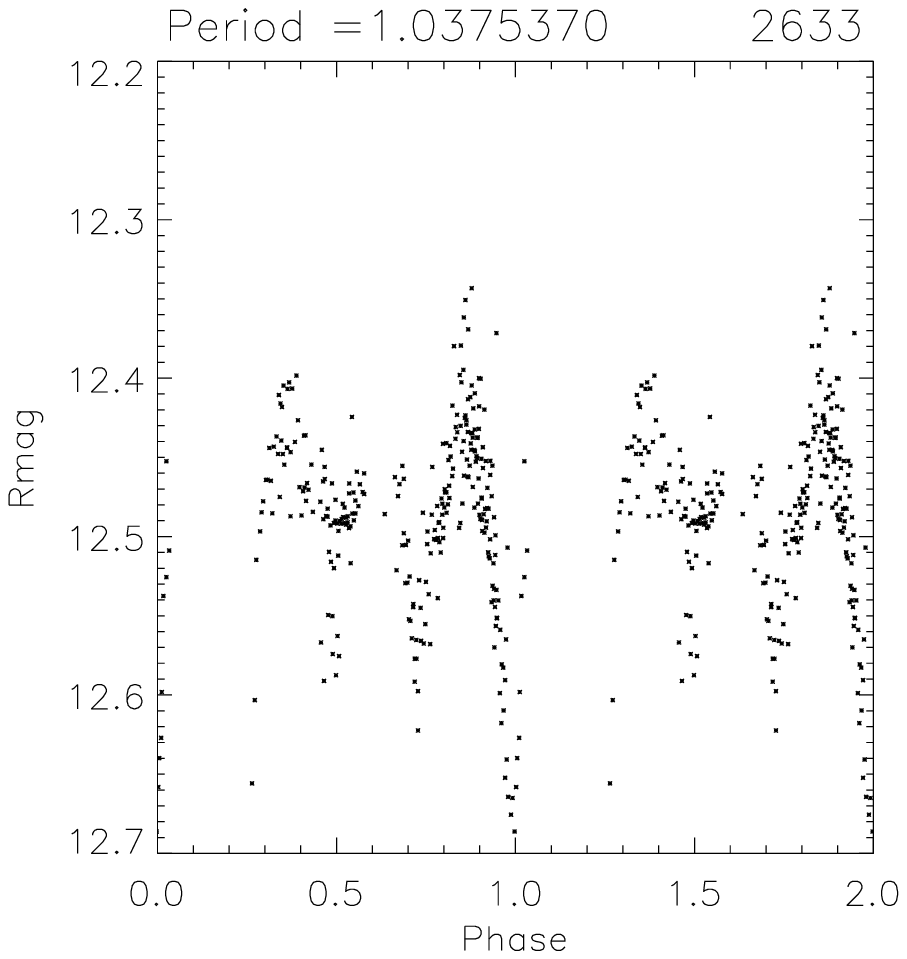}
\includegraphics[scale=.45]{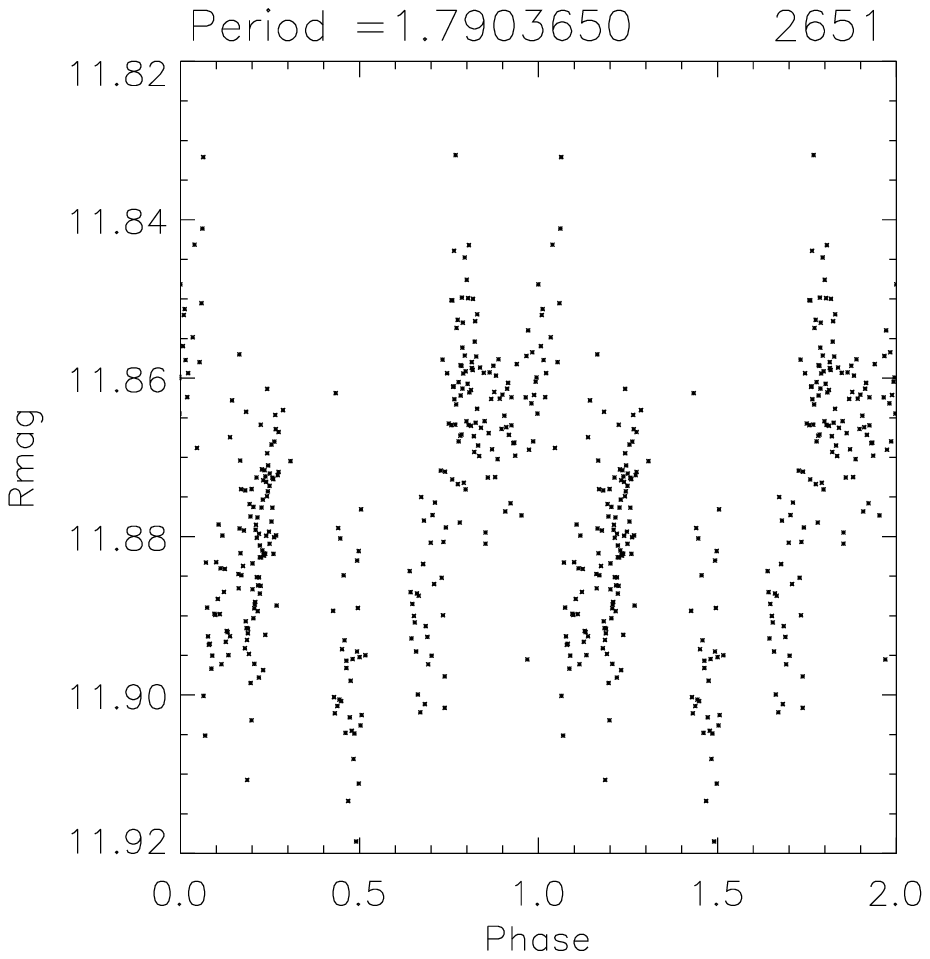}
\includegraphics[scale=.45]{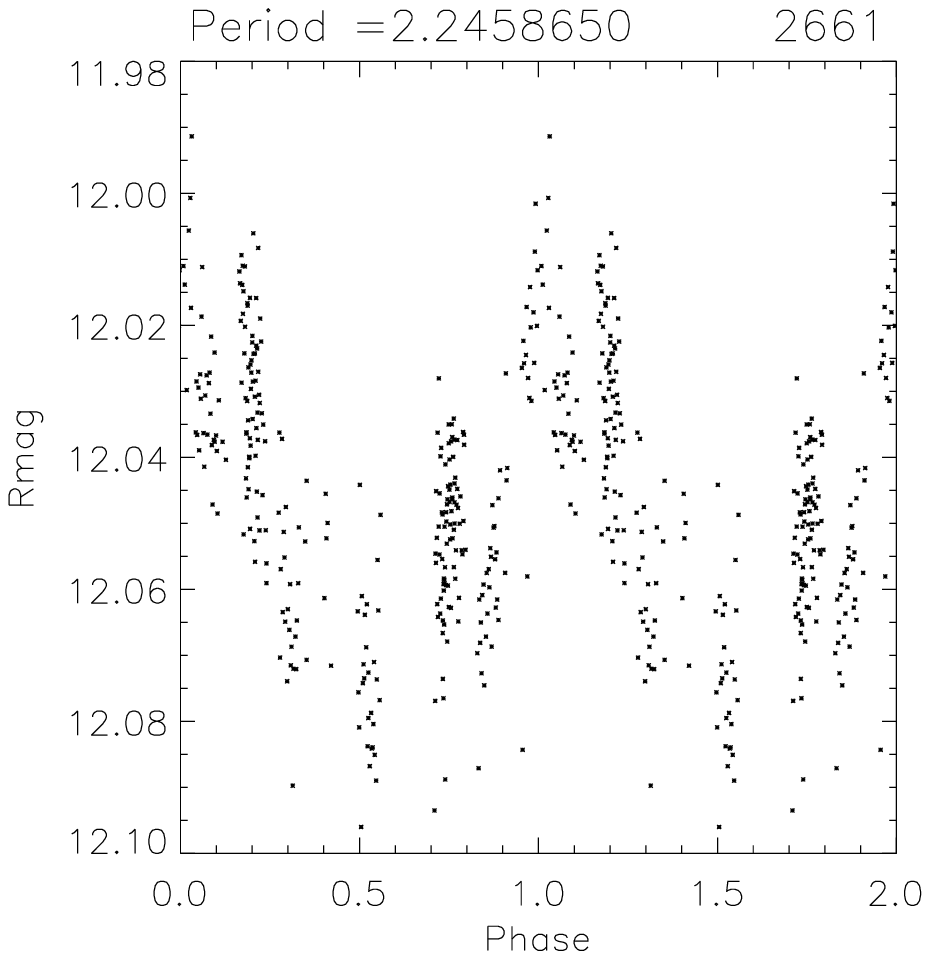}
\includegraphics[scale=.45]{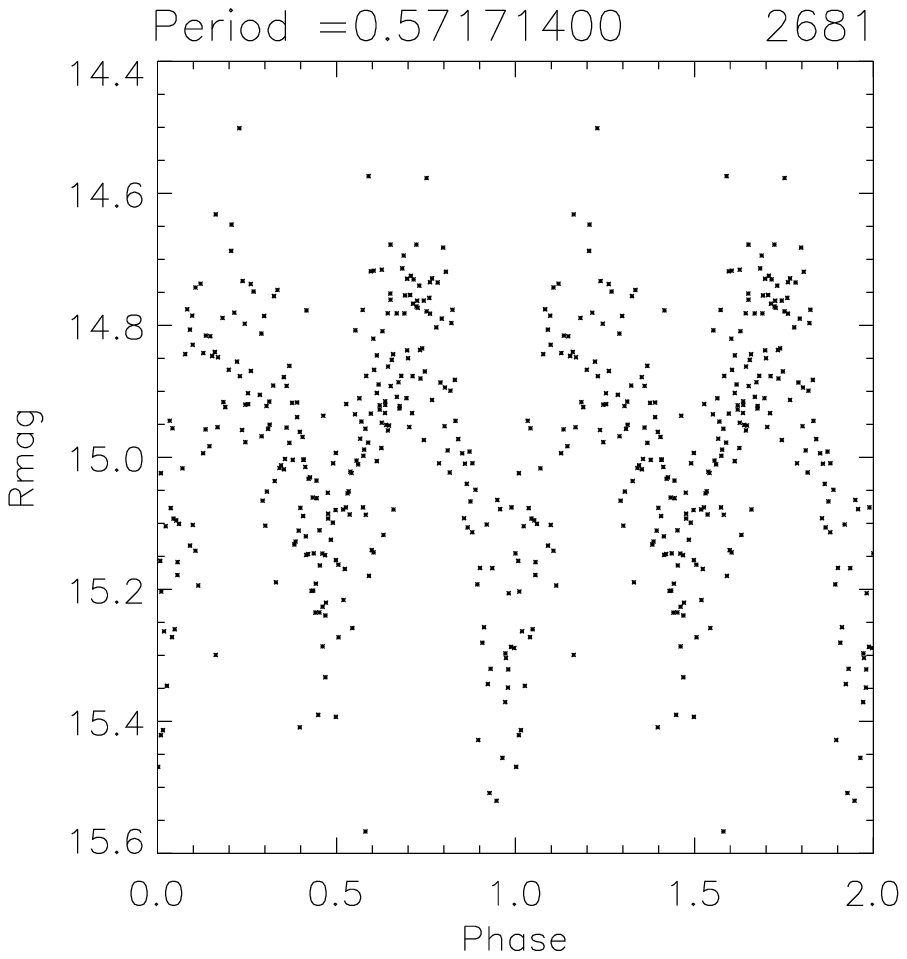}
\includegraphics[scale=.45]{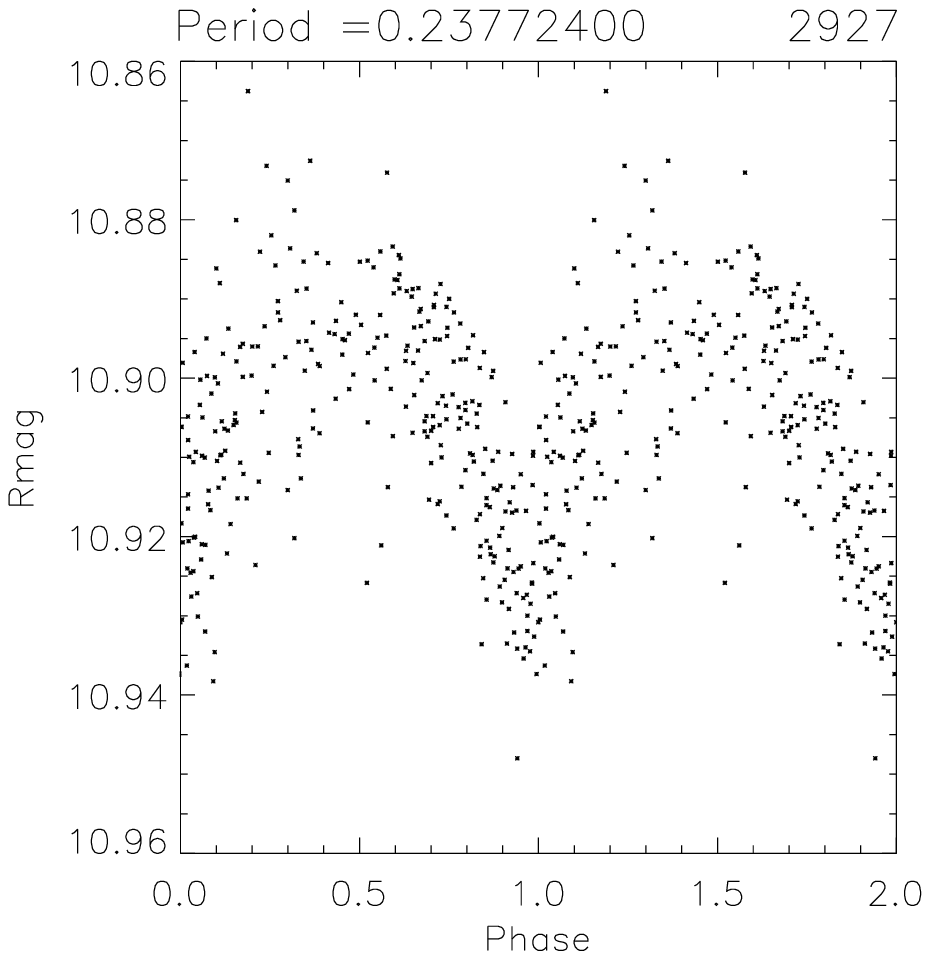}
\includegraphics[scale=.45]{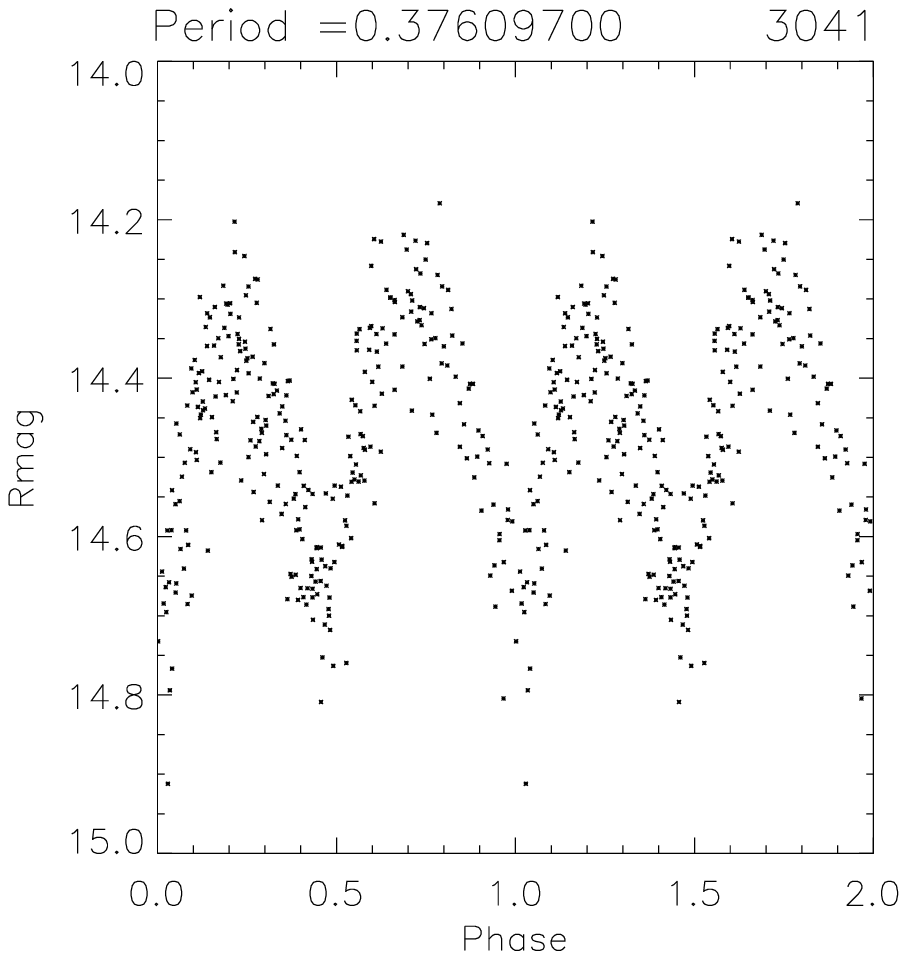}
\includegraphics[scale=.45]{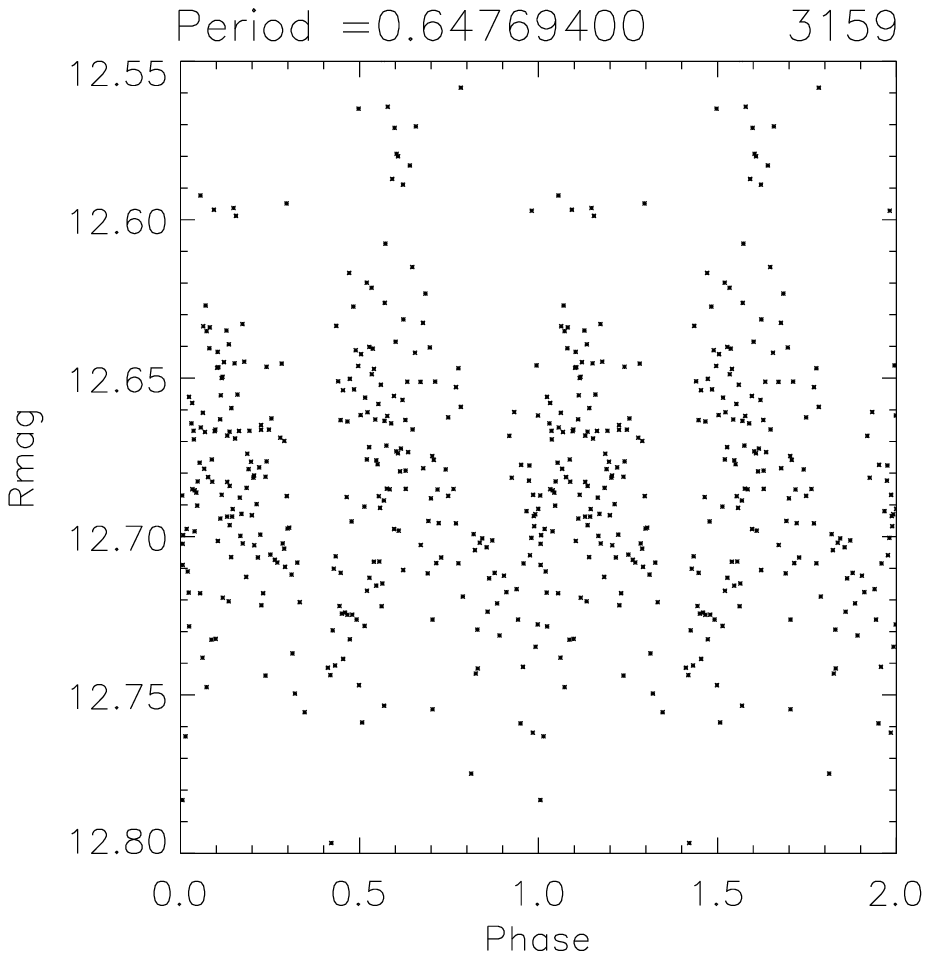}
\includegraphics[scale=.45]{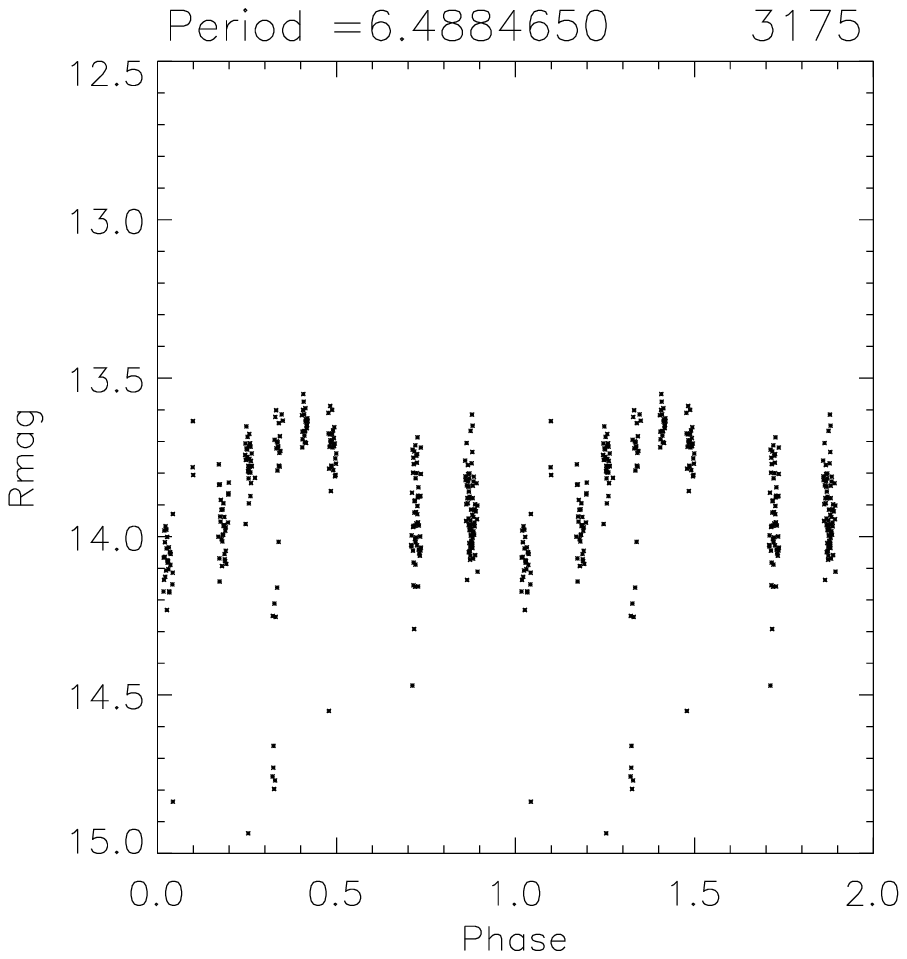}
\includegraphics[scale=.45]{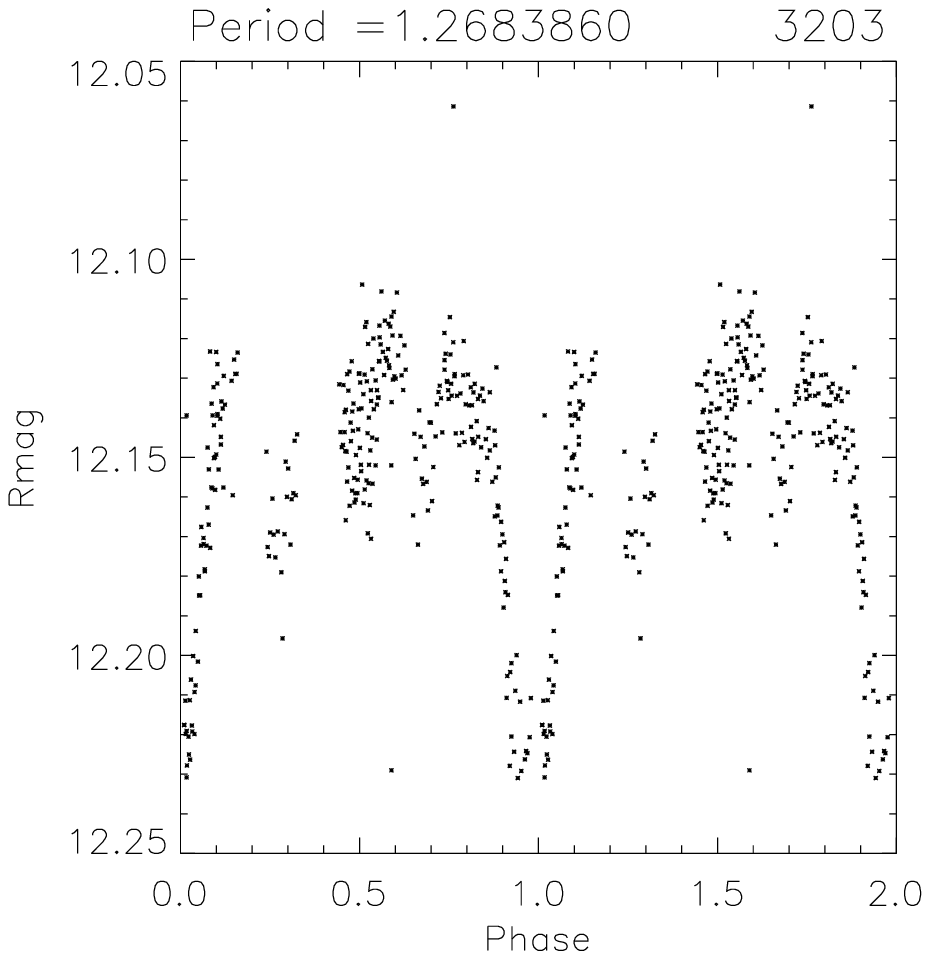}
\end{figure}
\clearpage
\begin{figure}[ht]
\includegraphics[scale=.45]{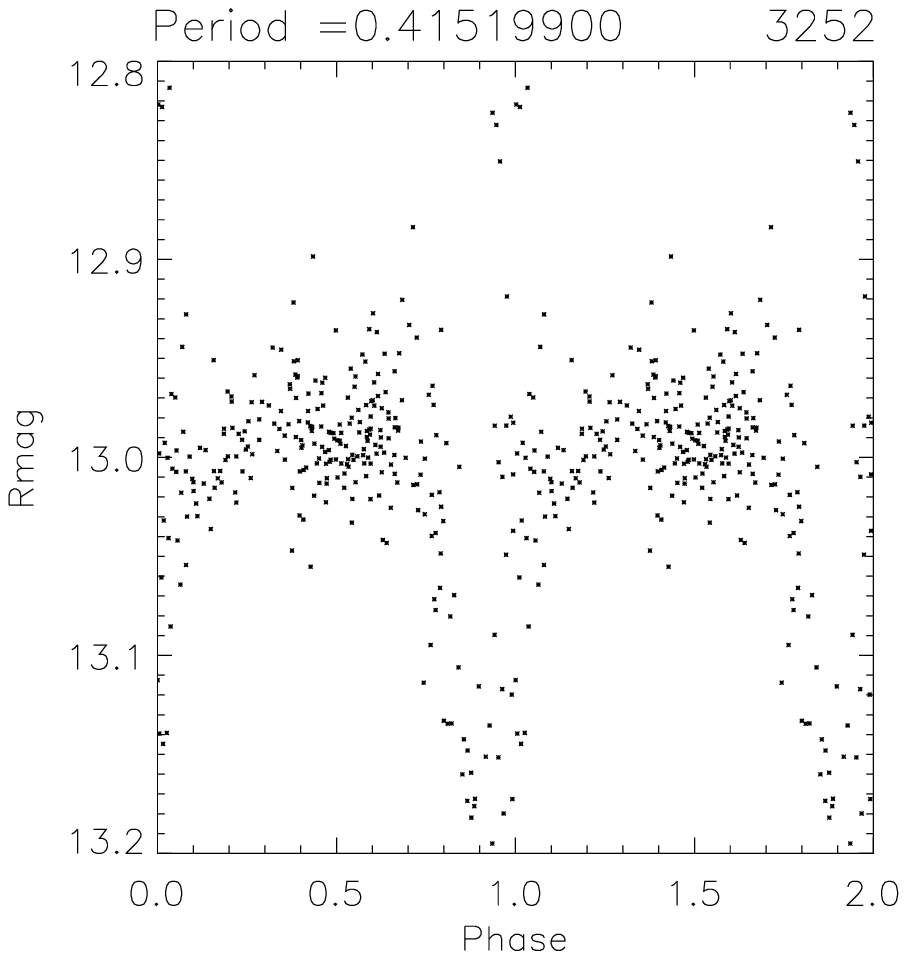}
\includegraphics[scale=.45]{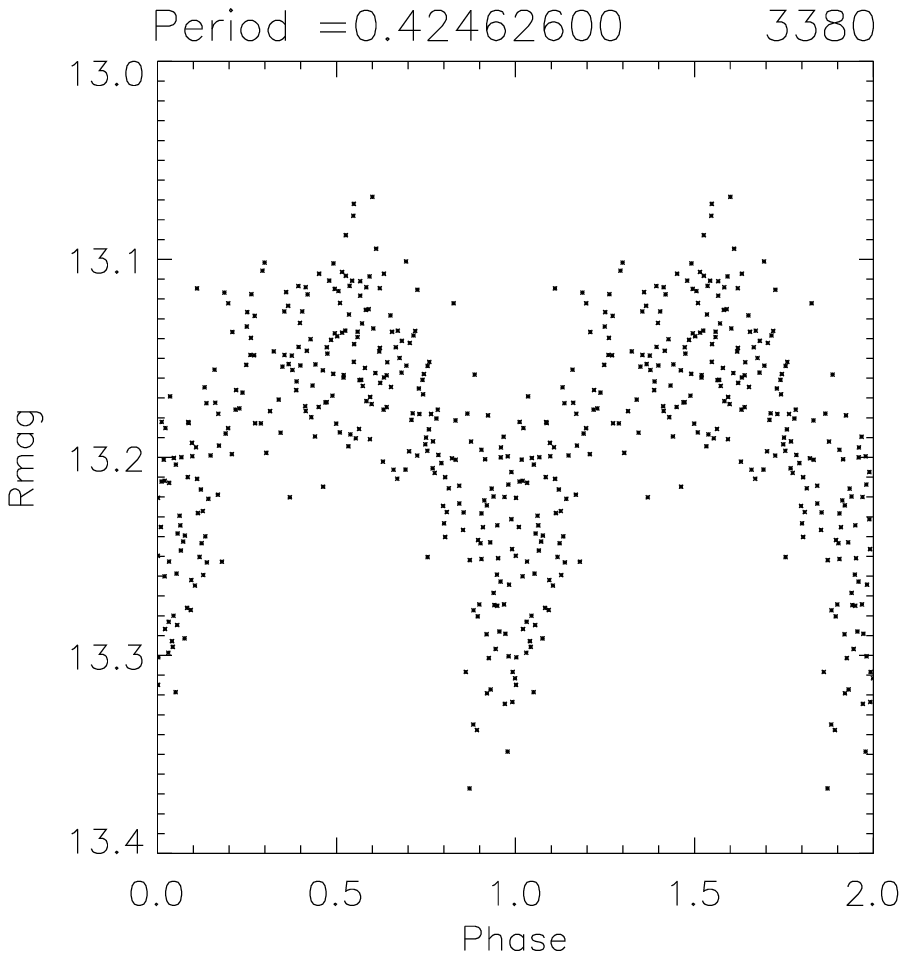}
\includegraphics[scale=.45]{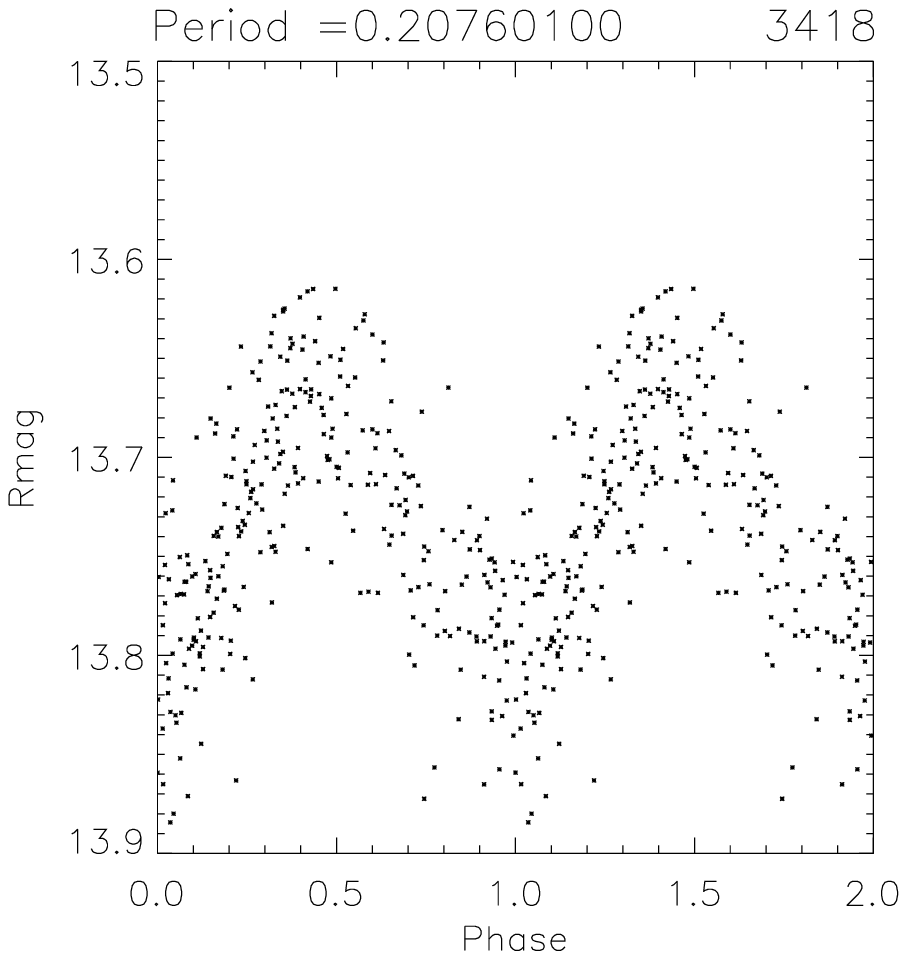}
\includegraphics[scale=.45]{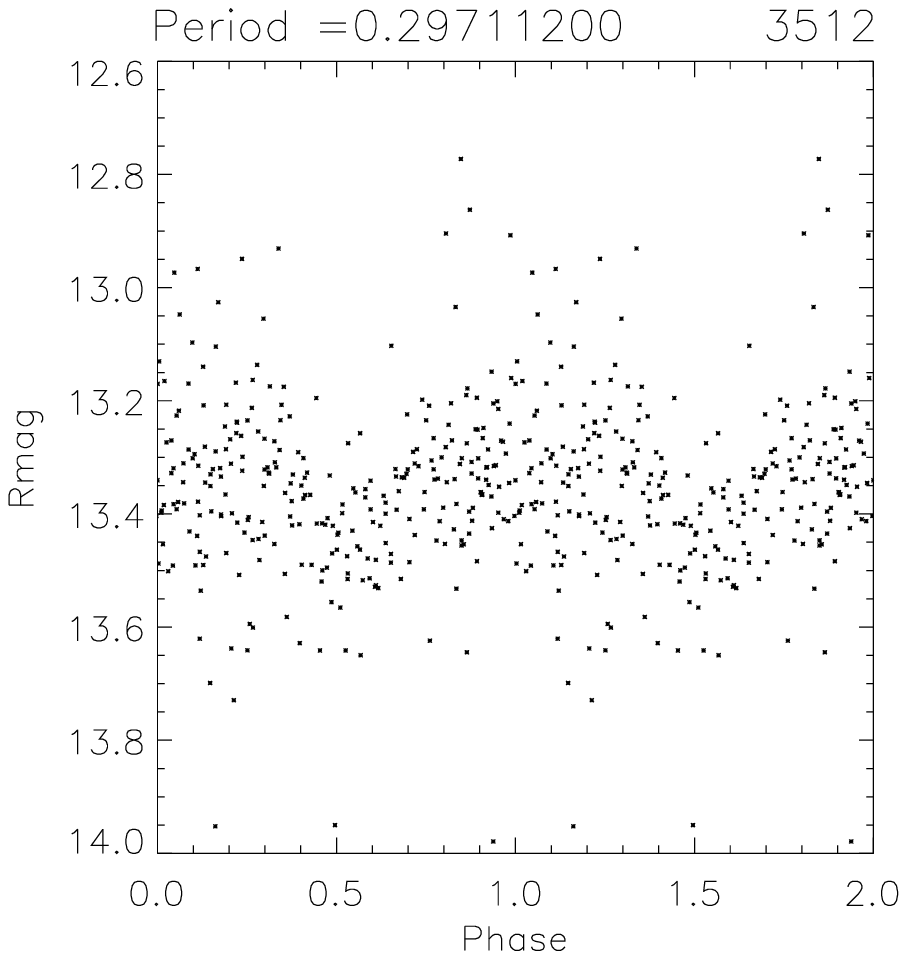}
\includegraphics[scale=.45]{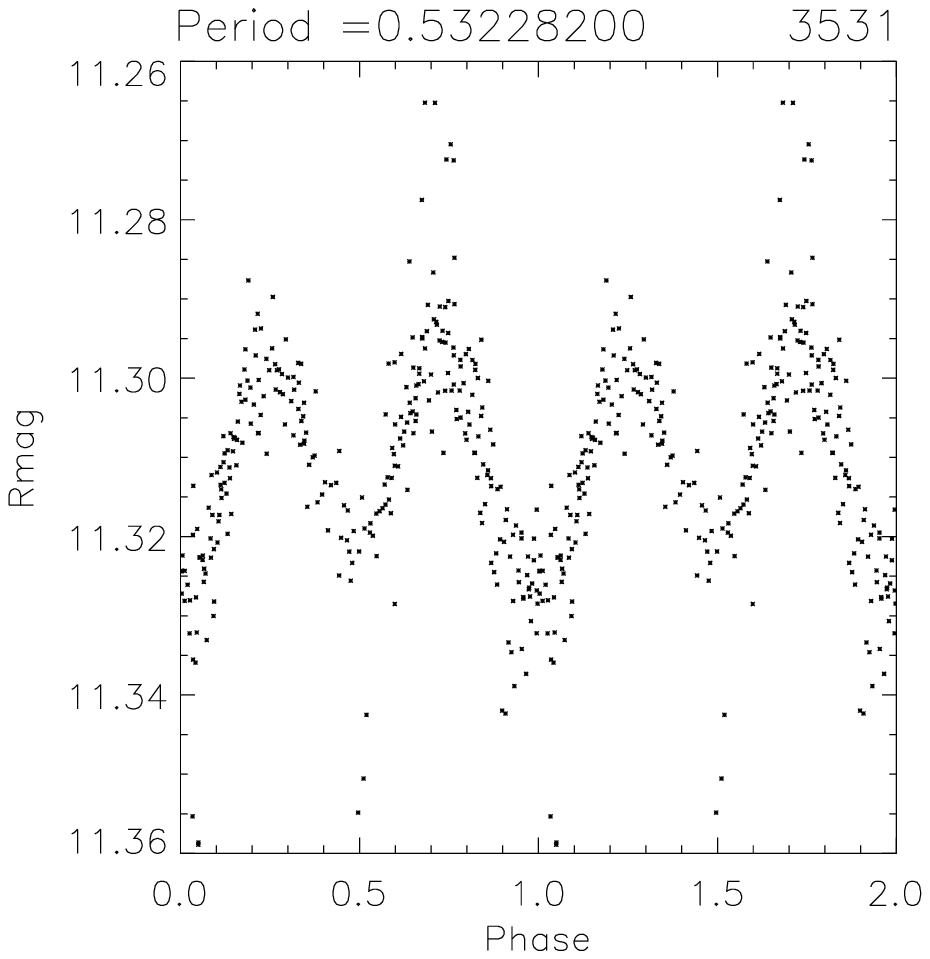}
\includegraphics[scale=.45]{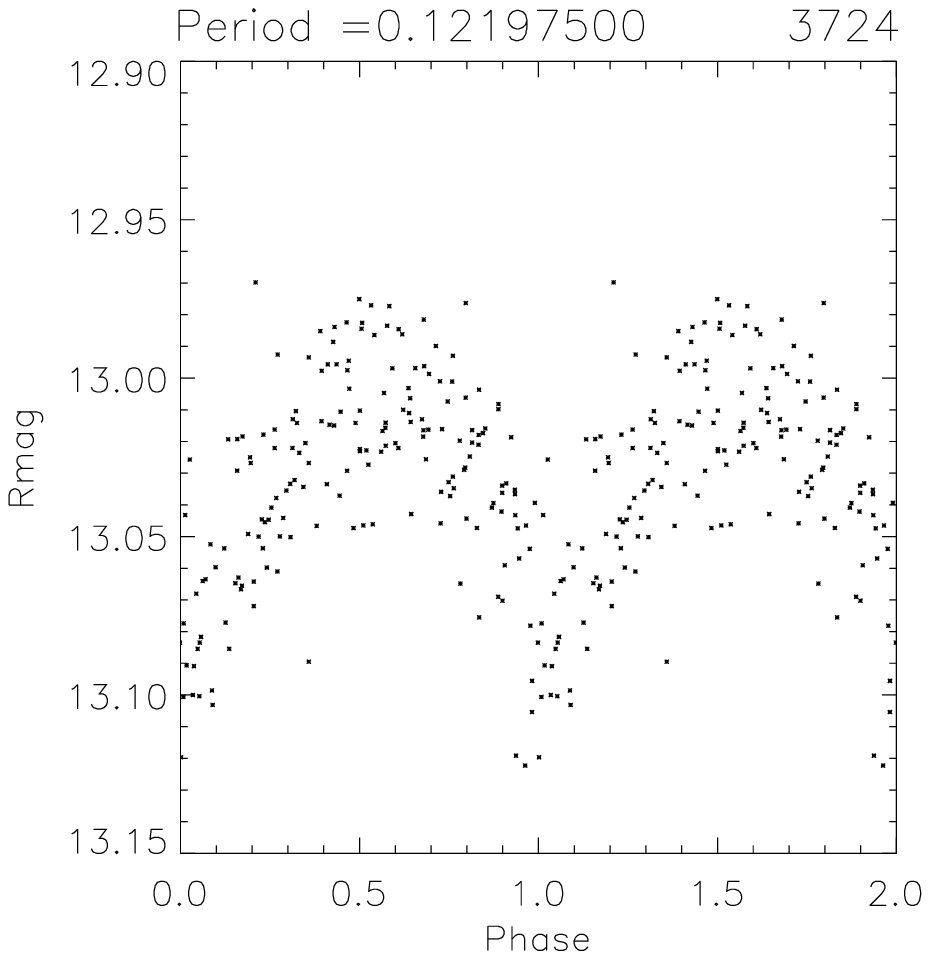}
\caption{Phased lightcurves of periodic variables. The BEST-ID is shown in the upper right corner of each graph. \label{variables}}
\end{figure}

\clearpage

\begin{deluxetable}{ccccccc}
\tablecaption{Equivalent Widths}
\tablewidth{0pt}
\tablecaption{Periodic variable stars detected. Magnitudes are based on calibration against USNO catalogue only. IDs marked with a star are within CoRoT field of view.}
\tablehead{
\colhead{BEST ID} & 
\colhead{$\alpha$(J2000)} &
\colhead{$\delta$(J2000)} & 
\colhead{Period(days)} &
\colhead{Mean mag} &
\colhead{Amplitude(mag)} &
\colhead{Type}
}    
\startdata
    1101 & 06$^{h}$43$^{m}$20$^{s}$&-3$^\circ$ 27' 25.5''&1.265 & 13.10 & 0.20 & ELL \nl
    1104 & 06$^{h}$50$^{m}$14$^{s}$&-3$^\circ$ 23' 21.0''&0.576 & 13.72 & 0.19 & CEP \nl
    1180 & 06$^{h}$50$^{m}$48$^{s}$&-3$^\circ$ 18' 11.3''&1.183 & 12.08 & 0.07 & ELL \nl
    1187 & 06$^{h}$46$^{m}$26$^{s}$&-3$^\circ$ 19' 42.7''&0.442 & 13.82 & 0.40 & EB  \nl
    1194 & 06$^{h}$45$^{m}$46$^{s}$&-3$^\circ$ 19' 51.7''&1.825 & 10.47 & 0.10 & ?  \nl
    1197 & 06$^{h}$46$^{m}$25$^{s}$&-3$^\circ$ 19' 22.2''&0.569 & 13.34 & 0.21 & CEP \nl
    1234 & 06$^{h}$42$^{m}$20$^{s}$&-3$^\circ$ 18' 54.3''&0.579 & 12.88 & 0.25 & CEP \nl
    1239 & 06$^{h}$42$^{m}$19$^{s}$&-3$^\circ$ 18' 25.9''&1.377 & 12.81 & 0.27 & CEP \nl
    1318 & 06$^{h}$50$^{m}$52$^{s}$&-3$^\circ$ 09' 39.9''&0.357 & 13.53 & 0.29 & EW  \nl
    1334 & 06$^{h}$49$^{m}$39$^{s}$&-3$^\circ$ 08' 54.6''&0.777 & 12.94 & 0.60 & EB  \nl
    1340 & 06$^{h}$49$^{m}$26$^{s}$&-3$^\circ$ 08' 34.8''&0.205 & 12.98 & 0.28 & EW  \nl
    1549* & 06$^{h}$47$^{m}$10$^{s}$&-2$^\circ$ 57'  0.7''&1.105 & 12.85 & 0.30 & EB  \nl
    1554* & 06$^{h}$47$^{m}$09$^{s}$&-2$^\circ$ 56' 52.0''&0.382 & 12.90 & 0.23 & EW  \nl
    1566 & 06$^{h}$44$^{m}$49$^{s}$&-2$^\circ$ 57' 43.4''&0.131 & 13.75 & 0.26 & EW  \nl
    1711 & 06$^{h}$41$^{m}$01$^{s}$&-2$^\circ$ 51' 17.2''&0.247 & 13.02 & 0.17 & EW  \nl
    1848 & 06$^{h}$44$^{m}$20$^{s}$&-2$^\circ$ 41'  0.2''&0.175 & 14.02 & 0.30 & EW  \nl
    1853 & 06$^{h}$44$^{m}$22$^{s}$&-2$^\circ$ 40' 47.7''&0.175 & 13.72 & 0.20 & EW \nl
    1858 & 06$^{h}$43$^{m}$07$^{s}$&-2$^\circ$ 41' 13.2''&0.638 & 12.10 & 0.30 & EA \tablebreak
    1865 & 06$^{h}$43$^{m}$07$^{s}$&-2$^\circ$ 40' 40.0''&3.534 & 12.08 & 0.21 & EB \nl
    1869 & 06$^{h}$44$^{m}$10$^{s}$&-2$^\circ$ 39' 45.6''&0.415 & 12.69 & 0.18 & EW \nl
    1881 & 06$^{h}$44$^{m}$13$^{s}$&-2$^\circ$ 39' 11.4''&0.208 & 13.38 & 0.32 & EW \nl
    1898 & 06$^{h}$51$^{m}$39$^{s}$&-2$^\circ$ 33' 33.6''&0.958 & 11.07 & 0.07 & ? \nl
    1990* & 06$^{h}$48$^{m}$07$^{s}$&-2$^\circ$ 29' 55.0''&0.233 & 11.94 & 0.06 & EA \nl
    1993* & 06$^{h}$51$^{m}$46$^{s}$&-2$^\circ$ 27' 43.9''&2.709 & 11.35 & 0.15 & ELL \nl
    2083* & 06$^{h}$46$^{m}$18$^{s}$&-2$^\circ$ 26'  2.4''&0.284 & 11.68 & 0.04 & DSCT \nl
    2091* & 06$^{h}$50$^{m}$41$^{s}$&-2$^\circ$ 22' 51.8''&0.169 & 12.40 & 0.07 & DSCT \nl
    2109 & 06$^{h}$50$^{m}$39$^{s}$&-2$^\circ$ 22'  6.7''&0.256 & 10.66 & 0.60 & EW \nl
    2141 & 06$^{h}$40$^{m}$39$^{s}$&-2$^\circ$ 26' 18.9''&0.239 & 12.68 & 0.60 & DSCT \nl
    2158 & 06$^{h}$46$^{m}$14$^{s}$&-2$^\circ$ 22' 30.8''&0.120 & 12.91 & 0.50 & EW \nl
    2325* & 06$^{h}$45$^{m}$27$^{s}$&-2$^\circ$ 13' 47.6''&0.255 & 13.59 & 0.26 & EW \nl
    2349* & 06$^{h}$49$^{m}$22$^{s}$&-2$^\circ$ 10' 15.6''&0.452 & 11.16 & 0.40 & EB \nl
    2354* & 06$^{h}$49$^{m}$24$^{s}$&-2$^\circ$ 09' 45.8''&0.827 & 10.31 & 0.40 & EB \nl
    2390* & 06$^{h}$49$^{m}$21$^{s}$&-2$^\circ$ 06' 48.5''&0.874 & 11.60 & 0.50 & EB \nl
    %2396* & 06$^{h}$49$^{m}$20$^{s}$&-2$^\circ$ 06' 27.5''&0.873 & 11.50 & 0.50 & EB \nl
    2440 & 06$^{h}$49$^{m}$14$^{s}$&-2$^\circ$ 04' 04.4''&4.006 & 12.02 & 0.08 & CEP \nl
    2484* & 06$^{h}$49$^{m}$09$^{s}$&-2$^\circ$ 02' 06.4''&0.213 & 14.10 & 0.22 & DSCT \nl
    2556* & 06$^{h}$45$^{m}$55$^{s}$&-1$^\circ$ 59' 54.4''&0.193 & 14.26 & 0.30 & DSCT \nl
    2628* & 06$^{h}$45$^{m}$49$^{s}$&-1$^\circ$ 55' 57.0''&0.518 & 11.39 & 0.26 & EB\nl
    2630* & 06$^{h}$49$^{m}$01$^{s}$&-1$^\circ$ 53' 42.8''&0.374 & 11.93 & 0.28 & EW \nl
    2632 & 06$^{h}$44$^{m}$04$^{s}$&-1$^\circ$ 56' 27.6''&2.251 & 12.43 & 0.05 & CEP \nl
    2633 & 06$^{h}$45$^{m}$50$^{s}$&-1$^\circ$ 55' 24.2''&1.038 & 12.49 & 0.30?& ? \nl
    2651 & 06$^{h}$44$^{m}$04$^{s}$&-1$^\circ$ 55' 05.0''&1.790 & 11.88 & 0.07 & CEP \nl
    2661 & 06$^{h}$44$^{m}$04$^{s}$&-1$^\circ$ 54' 33.5''&2.246 & 12.05 & 0.09 & CEP \nl
    2681* & 06$^{h}$49$^{m}$38$^{s}$&-1$^\circ$ 50' 18.9''&0.572 & 14.98 & 0.80 & EW \nl
    2927 & 06$^{h}$42$^{m}$33$^{s}$&-1$^\circ$ 40' 02.1''&0.238 & 10.91 & 0.05 & EW \nl
    3041 & 06$^{h}$41$^{m}$43$^{s}$&-1$^\circ$ 32' 25.8''&0.376 & 14.47 & 0.60 & EW \nl
    3159 & 06$^{h}$51$^{m}$38$^{s}$&-1$^\circ$ 21' 01.4''&0.648 & 12.69 & 0.14 & CEP \nl
    3175* & 06$^{h}$51$^{m}$36$^{s}$&-1$^\circ$ 19' 51.1''&6.489 & 13.89 & 0.65 & CEP \nl
    3203* & 06$^{h}$44$^{m}$15$^{s}$&-1$^\circ$ 22' 34.7''&1.268 & 12.15 & 0.15 & EA \nl
    3252* & 06$^{h}$44$^{m}$35$^{s}$&-1$^\circ$ 19' 08.7''&0.415 & 13.00 & 0.24 & EA \nl
    3380 & 06$^{h}$42$^{m}$16$^{s}$&-1$^\circ$ 10' 50.9''&0.425 & 13.18 & 0.25 & EW \nl
    3418 & 06$^{h}$49$^{m}$56$^{s}$&-1$^\circ$ 02' 59.3''&0.208 & 13.74 & 0.25 & EW \nl
    3512 & 06$^{h}$40$^{m}$27$^{s}$&-1$^\circ$ 03' 38.3''&0.297 & 13.35 & 0.50 & DSCT \nl
    3531 & 06$^{h}$42$^{m}$46$^{s}$&-1$^\circ$ 01' 09.7''&0.532 & 11.31 & 0.04 & EB \nl    
    3724* & 06$^{h}$45$^{m}$39$^{s}$&-0$^\circ$ 47' 33.8''&0.122 & 13.03 & 0.03 & EW \nl
    \enddata
\end{deluxetable}

\end{document}